\catcode`\@=11
{\count255=\time\divide\count255 by 60 \xdef\hourmin{\number\count255}
        \multiply\count255 by-60\advance\count255 by\time
   \xdef\hourmin{\hourmin:\ifnum\count255<10 0\fi\the\count255}}
\def\ps@draft{\let\@mkboth\@gobbletwo
    \def\@oddhead{}
    \def\@oddfoot
       {\hbox to 7 cm{$\scriptstyle Draft\ version:\ \draftdate$
       \hfil}
       \hskip -7cm\hfil\rm\thepage \hfil}
    \def\@evenhead{}\let\@evenfoot\@oddfoot}
\catcode`\@=12
\def\draftdate{\number\month/\number\day/\number\year\ \ \ \hourmin }

\documentstyle[12pt]{article}
\newcommand{\be}{\begin{eqnarray}}
\newcommand{\en}{\end{eqnarray}}
\newcommand{\non}{\nonumber}
\newcommand{\no}{\noindent}
\newcommand{\vs}{\vskip}
\newcommand{\hs}{\hspace}

\newcommand{\D}{\dagger}
\newcommand{\ef}{\`{e}}
\newcommand{\bC}{{\bf C}}
\newcommand{\p}{\partial}

\begin{document}
\
\vskip 2 cm

\no NON-COMPACT WZW CONFORMAL FIELD THEORIES\footnote{extended
version of lectures read at the Summer Institute ``New Symmetry
Principles in Quantum Field Theory'', Carg\ef se,
July 16-27, 1991}$^,$\footnote{based on joint work in progress
with Antti Kupiainen; these notes cover also some preliminary material
far from being completely understood, reflecting there the present
author's point of view which has been changing with time and has
not yet reached the final form}
\vskip 0.8 cm

\no\hs{2.2 cm}Krzysztof Gaw\c{e}dzki
\vs 0.3 cm

\no\hs{2.2 cm}C.N.R.S., I.H.E.S.
\vs 0.3 cm

\no\hs{2.2 cm}91440 Bures-sur-Yvette, France
\vs 0.7 cm

\no ABSTRACT
\vs 0.2 cm

We discuss non-compact WZW sigma models, especially the ones with
symmetric space $H^{\bf C}/H$ as the target, for $H$ a compact
Lie group. They offer examples of non-rational conformal field
theories. We remind their relation to the compact WZW models
but stress their distinctive features like the continuous
spectrum of conformal weights, diverging partition functions
and the presence of two types of operators analogous to the
local and non-local insertions recently discussed in the
Liouville theory. Gauging non-compact abelian subgroups of
$H^{\bf C}$ leads to non-rational coset theories. In
particular, gauging one-parameter boosts in the $SL(2,\bC)/SU(2)$
model gives an alternative, explicitly stable
construction of a conformal sigma
model with the  euclidean 2D black hole target.
We compute the (regularized) toroidal partition function
and discuss the spectrum of the theory. A comparison is made with
more standard approach based on the $U(1)$ coset
of the $SU(1,1)$ WZW theory where stability is not evident
but where unitarity becomes more transparent.
\vs 0.6 cm

\no 1.\ \ INTRODUCTION
\vs 0.3 cm

The four years which passed since the previous Carg\ef se
Institute of the series have brought a marked progress in the
understanding of rational Conformal Field Theories (CFT's),
a class of 2D massless quantum field models, see e.g. \cite{ms}.
The simplest of those theories is the free field with values
in a circle of rational radius,
more complicated examples are provided by the
Wess-Zumino-Witten (WZW) sigma models with a general
compact Lie group $G$ as the target \cite{w1},\cite{kz}\cite{gw} or by the
coset theories obtained by gauging a subgroup of $G$
in the WZW theory \cite{gko},\cite{bfo'r}. The characteristic property of
the rational CFT's is
\vs 0.1 cm

1. decomposition of the euclidean Green
functions into a finite sum of products of holomorphic and
antiholomorphic ``conformal blocks''.
\vs 0.1 cm

\no This is accompanied by other simplifying features,
which are more or less special from the point of view
of the general quantum field theory, like
\vs 0.1 cm

2. discreteness of finite volume energy spectrum,
\vs 0.1 cm

3. one-to-one correspondence between states and operators,
\vs 0.1 cm

4. simple structure of the operator product expansions,
\vs 0.1 cm

5. finiteness of the partition functions,
\vs 0.1 cm

6. simple factorization properties.
\vs 0.2 cm

\no Our knowledge about the rational WZW and coset
theories seems rather satisfactory today (although one
might argue that a subtle cleaning of some fine
mathematical points remains to be done; also the basic
problem of classification of the  rational CFT's has
not been solved). It contains the exact solution for
the spectrum and for the low genus Green functions (see e.g. \cite{quad}).
It seems then reasonable to go beyond the study of
relatively simple conformal theories where properties
1-6 hold, especially since examples of conformal models
without those properties appear rather naturally. The best
known instance is the Liouville theory describing the
conformal mode of 2D gravity \cite{polya}. It is in this model,
essential for the treatment of non-critical
string theory, where the new features related to the
failure of 1-6 where first discussed, see inspiring
lectures \cite{seib}.
\vs 0.1 cm

Mathematically, the passage from rational to irrational
CFT's involves a shift from purely algebraic treatment
to more analysis. The parallel might be the passage
from representation theory of compact Lie groups to the
non-compact case. Indeed, the canonical WZW example of
rational CFT is related to the representation theory
of loop groups of compact groups \cite{lg} and it is expected
that (largely non-existent, see however \cite{jk},\cite{dlp}) theory of
representations of loop groups of non-compact type
will underlie an interesting class of irrational CFT's.
The first candidates which come to mind are the
WZW theories with non-compact groups as targets.
These however, if quantized as in the compact case,
have unbounded below energy and, consequently,
no stability and no euclidean picture.
A possible solution is to
pass to their coset models where in some cases
one may expect to recover stability, see Sec. 5
below. In the present course, we shall start
instead from a different
series of non-rational CFT's which have bounded below
energy and stable euclidean picture but are
non-unitary. These are the WZW-type sigma models with
non-compact target spaces $H^{\bf C}/H$ where $H$ is a
compact (simple, connected, simply connected) Lie group.
We shall call them shortly $H^{\bf C}/H$ WZW models.
It should be stressed that, contrary to what the
name might suggest, this is a different class of
models than the coset $G/H$ theories in the CFT
sense. The latter are obtained by gauging
a subgroup $H\subset G$ (or more generally
$H\subset G_{\rm left}\times
G_{\rm right}$) in the group $G$ WZW model
and correspond rather to conformal sigma models with
orbit space of the left-right action of $H$ on $G$
as the target. To avoid the terminological
confusion, we shall label them as
$G\ {\rm mod}\ H$ coset theories\footnote{we
are fully aware that this arrogant attempt to change
accepted terminology is bound to be futile}.
In fact, a coset theory $G\ {\rm mod}\ H$
factorizes into the group $G$ WZW theory
times the $H^{\bf C}/H$ one, decoupled in the
planar topology, and, in general, coupled only via
zero modes. This is how the general $H^{\bf C}/H$
WZW theories manifested themselves
for the first time \cite{g/h},\cite{coset}. The $SL(2,\bC)/SU(2)$
model has been discussed earlier in \cite{haba}.
The Green functions of the $H^{\bf C}/H$ models
which appear in this context have also a 3D
interpretation: they compute the scalar product
of Schr\"{o}dinger picture
states of the 3D Chern-Simons \cite{w2},\cite{emss},\cite{quad}
field theory with gauge group $H$. In more geometric
terms, they give the hermitian structure
which pairs conformal blocks of the (rational)
group $H$ WZW model into its Green functions.
In this guise, the $H^{\bf C}/H$ theories may be
thought of as models dual to the ones with the
compact group $H$ as the target. All this is
briefly recalled in Sec. 2.
\vs 0.2 cm

In Sec. 3, we discuss free field
representations of the $H^{\bf C}/H$ WZW models
on the simplest example with $H=SU(2)$. We compute
explicitly (the finite part of) the partition
function of the model and discuss its spectrum
and the relation between the space of states
and the operators of the theory.
\vs 0.2 cm

The coset scenario for
producing new CFT's may also work in
the case of the $H^{\bf C}/H$
WZW models if one gauges out a non-compact abelian subgroup
$N\in H^\bC$ (the result will be called an $H^{\bf C}/H$
mod $N$ theory). In Sec. 4, using free field
representations, we show that the $SL(2,\bC)/SU(2)$
mod ${\bf R}$ model where $\bf R$ is embedded into
$SL(2,\bC)$ by $t\longmapsto{\rm e}^{t\sigma^3}$
gives a conformal sigma model with the recently
found \cite{w3},\cite{msw} 2D euclidean black hole as the target.
We discuss the partition functions and the spectrum
of this model. A comparison is made between the
$SL(2,\bC)/SU(2)$ mod ${\bf R}$ theory and the rational
parafermionic $SU(2)$ mod $U(1)$ model.
\vs 0.2 cm

Finally, in Sec. 5, we contrast our approach to the
black hole conformal sigma model with Witten's
original proposal \cite{w3} based on the $SU(1,1)$
mod $U(1)$ coset theory, see also \cite{dvv}-\cite{tsey}.
The free field calculation of the partition
functions of the black hole model may be also repeated
within Witten's scenario giving the same result but it requires
complex shifts and rotations of the fields in the
functional integral. One may reasonably expect that
both models coincide, the two approaches being complementary:
the $SL(2,\bC)/SU(2)$ mod ${\bf R}$ picture provides an explicitly
stable construction whereas the $SU(1,1)$ mod $U(1)$ approach
should be more useful for demonstrating unitarity of the
theory.
\vs 0.5 cm

\no 2.\ \ ORIGIN OF THE $H^{\bf C}/H$ WZW MODELS
\vs 0.3 cm

\no\underline{2.1.\ \ From the coset $G\ {\rm mod}\ H$
theories}
\vs 0.2 cm

Let us start by recalling the formulation of a coset
$G\ {\rm mod}\ H$
theory as a partially gauged, group $G$ WZW model (with compact $G$).
The basic fields on the closed Riemann surface
$\Sigma$ are
$G^\bC$-valued functions $g$ and gauge fields $A=
A_zdz+A_{\bar z}d{\bar z}$ with values in the
complexified Lie algebra ${\cal H}^\bC$ of a group $H\subset G$.
$H$ is supposed to be embedded into $G$ in two
possibly different ways: $\iota_{l,r} :H
\hookrightarrow G$.
We shall denote by ``tr'' the invariant form
on the Lie algebra ${\cal G}$ (${\cal H}$) of $G$ ($H$)
normalized to give $2$ as the length
squared of the longest roots.
We assume that via embeddings $\iota_{l,r}$, tr on
$\cal G$ induces a single invariant form on $\cal H$
equal $\eta$ times tr on $\cal H$. The euclidean action
of the coset model takes the form \cite{coset}
\be
kS(g,A)\ =\ kS(g)+\frac{_{ik}}{^{\pi}}\int\limits_\Sigma
{\rm tr}\hs{0.06 cm}[A_z^r(g^{-1}
\p_{\bar z}g)+(g\p_{z}g^{-1})A_{\bar z}^l
\non\\
+i{\rm Ad}_g(A_z^r)A_{\bar z}^l-i\eta A_z
A_{\bar z}\hs{0.07 cm}d^2z
\en
\no where the superscripts ``$l,r$'' refer to the
embeddings of $H$ into $G$. $kS(g)$ is the pure WZW action
\be
S(g)\ =\ -\frac{_1}{^{2\pi}}\int\limits_\Sigma{\rm tr}
\hs{0.06 cm}(g^{-1}\p_zg)(g^{-1}\p_{\bar z}g)
\hs{0.07 cm}d^2z
+\frac{_1}{^{24\pi i}}\int\limits_\Sigma
d^{-1}{\rm tr}\hs{0.06 cm}(g^{-1}dg)^{\wedge 3}
\en
\no where we have used a shorthand notation for the Wess-Zumino
topological term \cite{w1}. Coupling constant $k$ (``level'') is a
positive integer. Under the complex ($H^\bC$-valued) chiral
gauge transformations
\be
g\ \longmapsto\ h_1^lgh_2^{r\dagger}\non
\en
\vs -1.1 cm
\be
A_{\bar z}\ \longmapsto\ ^{h_1}\hs{-0.1 cm}A_{\bar z}
\equiv{\rm Ad}_{h_1}
(A_{\bar z})-ih_1\p_{\bar z}h_1^{-1}\ ,\non
\en
\vs -1.1 cm
\be
A_z\ \longmapsto\ ^{h_2}\hs{-0.1 cm}A_z\equiv
{\rm Ad}_{{h_2^{\dagger}}^{-1}}(A_z)
-i{h_2^{\dagger}}^{-1}\p_zh_2^\dagger\ ,\non
\en
\no action (1) transforms like follows:
\be
S(h_1^lgh_2^{r\dagger},\hs{0.04 cm}^{h_2}\hs{-0.13 cm}A_zdz+
\hs{0.04 cm}^{h_1}\hs{-0.16 cm}A_{\bar z}d\bar z)\ =\ S(g,A)
+\eta S(h_1h_2^\dagger,\hs{0.04 cm}^{h_2}\hs{-0.13 cm}A_zdz
+\hs{0.04 cm}^{h_1}\hs{-0.16 cm}A_{\bar z}d\bar z)\ .
\en
\no In particular, it is invariant under the unitary
gauge transformations with $H$-valued $h_1=h_2=h$ .
\vs 0.2 cm

The Green functions of the coset model are formally
given by the functional integral
\be
\int\ -\
{\rm e}^{-kS(g,A)}\hs{0.05 cm}Dg\hs{0.05 cm}DA
\en
\no over $G$-valued fields $g$ and real (i.e. $\cal H$-valued)
gauge fields $A$. As the insertion, we should take an
expression invariant under the unitary gauge transformations.
An example is provided by
\be
\prod\limits_{\alpha}{\rm tr}_{R_{\alpha}}\hs{0.03 cm}g(\xi_{\alpha})n_{\alpha}
\en
\no where ``tr$_R$'' stands for the trace
in representation $R$ of $G$ (in vector space $V_R$)
and $n_{\alpha}\in G$ satisfy
\be
u^ln_{\alpha}u^{r\dagger}=n_{\alpha}
\en
\no for $u\in H$. For example, if $u^l=u^r$, we may take $n_\alpha=1$.
\vs 0.2 cm

On the Riemann sphere, we may parametrize real gauge fields
$A$ by $H^\bC$-valued gauge transformations by
putting $A_{\bar z}(h)=h^{-1}\p_{\bar z}h$. Action (1)
becomes then
\be
kS(g,A(h))\ =\ kS(h^lgh^{r\dagger})-\eta kS(hh^\dagger)\ .
\en
\no The Jacobian of the change of variables is (we ignore the
zero modes for the moment)
\be
\frac{\p(A(h))}{\p(h)}\
=\ {\rm det}(\bar{\p}_h^*\bar{\p}_h)
=\ {\rm e}^{2h^\vee{}S(hh^\dagger)}\hs{0.05 cm}
({\rm det}(-\Delta))^{{\rm dim}H}
\en
\no where $\bar{\p}_h=d\bar{z}
(\p_{\bar z}+{\rm ad}_{A_{\bar z}(h)})$, $h^\vee{}$ is the
dual Coxeter number of $H$ and $\Delta$
is the scalar Laplacian. More exactly, the change of
variables $A\mapsto h$ gives the following
expression for the Green functions (4) with
insertion (5):
\be
C\int\bigg(\prod\limits_{\alpha}{\rm tr}_{R_{\alpha}}\hs{0.06
cm}g(\xi_{\alpha})
\hs{0.06 cm}(h^ln_{\alpha}h^{r\dagger})^{-1}(\xi_{\alpha})\bigg)\hs{0.1 cm}
{\rm e}^{-kS(g)}\hs{0.07 cm}
{\rm e}^{(\eta k+2h^\vee{})S(hh^\dagger)}
\hs{0.06 cm}Dg\hs{0.09 cm}
\delta(h(\xi_0))\hs{0.08 cm}Dh
\en
\no where $C=\left({\rm det}'(-\Delta)/{\rm area}
\right)^{{\rm dim}H}$
with the determinant without the zero mode contribution.
Expression (9) combines the Green functions of the
compact group $G$ WZW model
\be
\Gamma\ =\ \int(\bigotimes\limits_{\alpha}g_{R_{\alpha}}(\xi_{\alpha}))
\ {\rm e}^{-kS(g)}\hs{0.09 cm}Dg\ \in\ \bigotimes
\limits_{\alpha} {\rm End}\hs{0.08 cm}V_{R_{\alpha}}
\en
\no (where $g_R$ denotes the representation $R$
matrix of $g$) with those of a field theory
with fields $hh^\dagger$
\be
\int(<\Gamma,\otimes
(h^ln_{\alpha}h^{r\dagger})^{-1}_{R_{\alpha}}(\xi_{\alpha})\hs{0.07 cm})>
\ {\rm e}^{\kappa S(hh^\dagger)}
\ \delta(hh^\dagger(\xi_0))\ D(hh^\dagger)\ .
\en
\no In the last expression $\Gamma$ may be any tensor
in $\otimes{\rm End}V_{R_{\alpha}}$ such that
\be
(\otimes \gamma^l_{R_{\alpha}})\Gamma(\otimes
\gamma^{r\dagger}_{R_{\alpha}})=\Gamma
\en
\no for $\gamma\in H^\bC$. This condition guarantees that
the integral is independent of
point $\xi_0$ in the $\delta$-function
in (11) fixing the global $H^\bC$ invariance. Green
functions (10) certainly satisfy condition (12). $<\cdot,\cdot>$
in (11) stands for the scalar
product induced from that of spaces $V_R$. Fields $hh^\dagger$
may be viewed as taking values in the non-compact symmetric
space $H^\bC/H$ and functional integral (11)
as defining the (euclidean) Green functions of the $H^\bC/H$
WZW theory (also in a general world-sheet
topology). The euclidean action $-\kappa S(hh^\dagger)$ of the
model is unambiguously defined\footnote{$H^\bC/H$
is topologically trivial} and real, non-negative \cite{coset}.
We shall see that it leads to functional integrals of type
(11) which are stable for any real $\kappa>h^\vee$. On the
other hand, the Minkowskian action is not real: the
Wess-Zumino term is purely imaginary so that we should not
expect the theory to be unitary. We shall return to these
issues below.
\vs 0.2 cm

On a higher genus Riemann surface a similar treatment
of the coset theory Green functions
produces again a combination of the $G$ and $H^\bC/H$
WZW Green functions but this time both twisted by coupling to
an external flat gauge field $A_{\rm flat}$ and the
result contains an integral over the
moduli of $A_{\rm flat}$ \cite{coset}, essentially coinciding
with the moduli of complex $H^{\bC}$-bundles.
\vs 0.4 cm

\no\underline{2.2.\ \ From the scalar product of the
Chern-Simons theory states}
\vs 0.2 cm

The Schr\"{o}dinger picture states of the 3D Chern-Simons
theory with gauge group $H$ on manifold $\Sigma\times{\bf R}$
and in the presence of the Wilson lines
$\{\xi_{\alpha}\}\times{\bf R}$ in representations $R_{\alpha}$
are functionals
\be
\psi:\ {\cal A}\ \longrightarrow\ \bigotimes\limits_{\alpha}
V_{R_{\alpha}}
\en
\no on space ${\cal A}$ of real gauge fields $A$ \cite{wzw}. $\cal A$
has a natural complex structure
obtained by identifying it with the
space of forms $A_{\bar z}d\bar z$. Functionals $\psi$
are required to be holomorphic and to transform covariantly
under the complex gauge transformations:
\be
\psi(^h\hs{-0.11 cm}A)\ =\ {\rm e}^{kS(h^{-1})+\pi^{-1}ik
\int{\rm tr}\hs{0.04 cm}(h^{-1}\p_zh)A_{\bar z}\hs{0.04 cm}
d^2z}\hs{0.04 cm}
\bigotimes\limits_{\alpha}h_{R_{\alpha}}(\xi_{\alpha})\ \psi(A)\ .
\en
\no $k$ this time denotes the
coupling constant of the Chern-Simons theory.
The space of states defined as above is
finite-dimensional. The scalar product of the states
is formally given by the functional integral
\be
\|\psi\|^2\ =\ \int <\psi(A),\psi(A)>\
{\rm e}^{\hs{0.04 cm}-\pi^{-1}k\int{\rm tr}\hs{0.04 cm}A_z
A_{\bar z}\hs{0.04 cm}d^2z}\ DA\ .
\en
\no On $\Sigma=\bC P^1$, upon the change of variables $A\mapsto h$,
eq.\hs{0.06 cm}(15) becomes
\be
\|\psi\|^2\
=\ \left({\rm det}'(-\Delta)/
{\rm area}\right)^{{\rm dim}H}\hs{5.2 cm}\non\\
\cdot\int <\psi(0)\otimes
\overline{\psi(0)}\hs{0.05 cm},\hs{0.05 cm}
\otimes (hh^\dagger)_{R_{\alpha}}^{-1}(\xi_{\alpha})>\
{\rm e}^{(k+2h^\vee{})S(hh^\dagger)}\ \delta(hh^\dagger
(\xi_0))\ D(hh^\dagger)
\en
\no which is a Green function of type (11) (for
$G=H$ and $n_{\alpha}\equiv 1$).
\vs 0.3 cm

\no\underline{2.3.\ \ From\ the\ hermitian\ structure\ coupling\
conformal\ blocks\ of\ the\ group\ $H$\ WZW}
\break\underline{theory}
\vs 0.2 cm

Green functions of the group $H$ WZW model in an
external $\cal H$-valued field $A$
\be
\int(\bigotimes\limits_{\alpha}g(\xi_{\alpha})_{R_{\alpha}})
\ {\rm e}^{-kS(g,A)}\hs{0.09 cm}Dg
\en
\no can be expressed as
\be
\sum\limits_{a,b}\Omega^{ab}\hs{0.07 cm}
\psi_{a}(A)\otimes\overline{\psi_{b}(A)}\
{\rm e}^{\hs{0.04 cm}-\pi^{-1}k\int\hs{0.04 cm}{\rm tr}\hs{0.04 cm}
A_zA_{\bar z}\hs{0.04 cm}d^2z}
\en
\no where $(\psi_a)$ is a basis of the
Chern-Simons states considered above and the inverse matrix
\be
(\Omega^{-1})_{ab}\ =\ (\psi_a,\psi_b)
\en
\no in the scalar product of (15), see \cite{quad},\cite{ccft}. In the planar
or toroidal geometry, the dependence of the basis vectors
$\psi_a$ on the insertion points and the complex
structure may be chosen analytic and such that the scalar
products $(\psi_a,\psi_b)$ remain constant.
Expression (18) gives then the
decomposition of the Green functions into sum of combinations
of conformal blocks demonstrating the rational character
of the WZW theories with compact targets. As we see,
scalar product (15) given by the Green functions
of the $H^\bC/H$ theory determines the way the conformal
blocks of group $H$ WZW theory are put together to build
the complete Green functions.
\vs 0.02 cm

The WZW theories with targets $H$ and $H^\bC/H$ may be
considered as dual to each other. An elegant way to
express this duality is to consider the coset
$H$ mod $H$ model. This is a topological theory
in the sense of \cite{w4}: its Green functions
\be
\int(\prod\limits_{\alpha}{\rm tr}_{R_{\alpha}}\hs{0.03 cm}g(\xi_{\alpha}))
\ {\rm e}^{-kS(g,A)}\hs{0.09 cm}Dg\hs{0.07 cm}DA
\en
\no are independent of the location of the insertions
and of the complex structure of the surface \cite{coset}.
Integrating representation (18) over gauge fields $A$,
one infers \cite{hv} that they are in fact equal to the
dimensions of the spaces of states $\psi$ known explicitly
due to \cite{ev}. On the
other hand, the coset Green functions factorize,
as we have seen, into a combination of products of
those of the group $H$ and of the symmetric space
$H^\bC/H$ WZW models. This is the precise expression
of the duality between both theories.
\vs 0.2 cm

In the planar case, the $H^{\bC}/H$ theory with Green functions (11)
may be also viewed as an analytic continuation of those of the $H$ theory
to negative levels. This relation becomes more complicated
on higher genera as, for example, a look into the respective
partition functions shows. It is not excluded, however, that
both models describe different aspects of the same structure
analytic in $k$.
\vs 0.5 cm

\no 3.\ \ FREE FIELD REPRESENTATION OF THE $H^\bC/H$ WZW THEORY
\vs 0.3 cm
\addtocounter{equation}{-20}

Functional integral (2.11) defining the Green functions
of the $H^\bC/H$ WZW theory may be computed by iterative
Gaussian integration. This was noticed in \cite{haba} for the $H=SU(2)$
case and was implemented in the present context and for general
$H$ in \cite{g/h},\cite{coset} for
the twisted toroidal partition function and in \cite{quad},\cite{fgk} for the
planar Green functions. One can also compute toroidal Green functions.
Free field representation
for the model on a surface of genus $>2$
is still an open problem. Below, we shall stick to the $SU(2)$ case,
for simplicity.
\vs 0.02 cm

Symmetric space $SL(2,\bC)/SU(2)$ coincides with the upper sheet
$H^{+}_3$ of 3D mass hyperboloid. Convenient global coordinate system
on $H^{+}_3$ is provided by the parametrization
\be
hh^\D\ =\ \left(\matrix{{\rm e}^{\phi}(1+|v|^2)^{1/2}&v\cr
\bar v&{\rm e}^{-\phi}(1+|v|^2)^{1/2}\cr}\right)
\en
\no with $\phi$ real and $v$
complex. The $SL(2,\bC)$-invariant measure on $H^{+}_3$\hs{0.04 cm},
\hs{0.05 cm}$d(hh^\D)=d\phi\hs{0.05 cm}d^2v$.
In coordinates (1),
\be
S(hh^\dagger)\ =\ -\frac{_1}{^{\pi}}\int[\hs{0.05 cm}
(\p_z\tilde\phi)(\p_{\bar z}\tilde\phi)+
(\p_z+\p_z\tilde\phi)\bar v\hs{0.08 cm}
(\p_{\bar z}+\p_{\bar z}\tilde\phi)v)\hs{0.05 cm}]\hs{0.07 cm}
d^2z
\en
\no where\footnote{it will become clear below why we use $\phi$ and not
$\tilde\phi$ in parametrizing $H^+_3$}
$\tilde\phi\equiv\phi-\frac{_1}{^2}{\rm log}(1+|v|^2)$.
We shall also need a gauged version of the action. If
we gauge the $U(1)$ group
embedded into $SU(2)$ asymmetrically by $\iota_l({\rm e}^{i\theta})
={\rm e}^{i\theta\sigma^3}$, $\iota_r({\rm e}^{i\theta})
={\rm e}^{-i\theta\sigma^3}$, then the transformation law
(2.3) implies, for $h_1={\rm e}^{\lambda\sigma^3}$ and
$h_2={\rm e}^{-\lambda\sigma^3}$, that
\be
S({\rm e}^{\lambda\sigma^3}\hs{-0.05 cm}g\hs{0.04 cm}
{\rm e}^{\lambda\sigma^3},
\hs{0.05 cm}A+id\lambda)
\ =\ S(g,A)
\en
\no for any $SL(2,\bC)$-valued $g$
(in particular for $g=hh^\dagger$)
and for any complex 1-form $A$.
Consequently, taking $A$ purely imaginary may be interpreted
as gauging of subgroup ${\bf R}\hookrightarrow
\{{\rm e}^{\lambda\sigma^3}\hs{0.08 cm}|\hs{0.08 cm}
\lambda\ {\rm real}\}$ in $SL(2,\bC)$ (which is the global
symmetry group of the $H_{3}^+$ WZW model). $\bf R$ corresponds
to the boosts in the third direction
under the standard relation between $SL(2,\bC)$ and
the Lorentz group. A direct computation gives
\be
S(hh^\dagger,\frac{_1}{^{2i}}A)\ =\ -\frac{_1}{^{\pi}}
\int[\hs{0.05 cm}
(\p_z\tilde\phi+A_z)(\p_{\bar z}\tilde\phi+A_{\bar z})
\non\\
+\hs{0.05 cm}(\p_z+\p_z\tilde\phi+A_z)\bar v\hs{0.08 cm}
(\p_{\bar z}+\p_{\bar z}\tilde\phi+A_{\bar z})v\hs{0.05 cm}]
\hs{0.08 cm}d^2z\ .
\en
\no Invariance (3) becomes obvious in (4)
since transformation $hh^\dagger\longmapsto
{\rm e}^{\lambda\sigma^3}hh^\dagger
{\rm e}^{\lambda\sigma^3}$ translates in coordinates (1)
into $(\phi,v)\longmapsto(\phi+2\lambda,v)$.
\vs 0.3 cm

\no\underline{3.1.\ \ Toroidal partition function}
\vs 0.2 cm

First, let us describe the calculation \cite{coset} of the twisted
partition function ${\cal Z}^{H^+_3}(\tau,U)$ of the $H^+_3$ WZW
theory on torus $T_\tau\equiv\bC/(2\pi{\bf Z}+2\pi\tau{\bf Z})$,
$\tau=\tau_1+i\tau_2,\ \tau_2>0$. It is given
by the functional integral:
\be
{\cal Z}^{H^+_3}(\tau,U)\ =\ \int{\rm e}^{\hs{0.04 cm}\kappa S(\gamma_U
hh^\D\gamma_U^\D)}\hs{0.08 cm}D(hh^\D)
\en
\no where $\gamma_U=\exp[-\frac{_1}{^{4\tau_2}}U(z-\bar z)
\sigma^3],\ U\equiv U_1+iU_2,$ satisfies
\be
\gamma_U(z+2\pi)=\gamma(z)\ \ {\rm and}\ \ \gamma_U(z+2\pi\tau)
={\rm e}^{-\pi iU\sigma^3}\gamma_U(z)\non
\en
\no and the action is extended to twisted field
configurations \cite{coset} by putting
\be
S(\gamma_Uhh^\dagger\gamma_U^\dagger)\ =
\ S(hh^\dagger\hs{0.05 cm},\hs{0.04 cm}\frac{_1}{^{4i}}
(\tau_2^{-1}\bar Udz+\tau_2^{-1}Ud\bar z))\ +\
\frac{_{\pi}}{^{\tau_2}}U_1^2\ .
\en
\no Using the explicit form (4) of the action, we obtain
\be
{\cal Z}^{H^+_3}(\tau,U)\ =\ {\rm e}^{\hs{0.03 cm}\pi\kappa\tau_2^{-1}
U_1^2}\hs{0.03 cm}\int{\rm e}^{\hs{0.03 cm}-\pi^{-1}\kappa
\hs{-0.03 cm}\int\hs{0.03 cm}(\p_z{\phi}
+\tau_2^{-1}\bar U/2)
(\p_{\bar z}{\phi}+
\tau_2^{-1}U/2)\hs{0.05 cm}d^2z}\ \non\\
\cdot\hs{0.1 cm}{\rm e}^{
\hs{0.03 cm}\int(\p_z+\p_z{\phi}+\tau_2^{-1}\bar U/2)
\bar v\hs{0.06 cm}(\p_{\bar z}+\p_{\bar z}{\phi}+
\tau_2^{-1}U/2)v\hs{0.04 cm}d^2z}
\hs{0.1 cm}D(hh^\D)\ .
\en
\no where we have shifted $\tilde\phi\mapsto\phi$.
The $v$-integral is gaussian and produces
\be
{\rm det}\hs{-0.04 cm}\left((\bar\p+\bar\p{\phi}
+\frac{_1}{^2}\tau_2^{-1}Ud\bar z)^{^*}\hs{0.04 cm}(\bar\p+\bar\p{\phi}
+\frac{_1}{^2}\tau_2^{-1}Ud\bar z)\right)^{-1}\hs{3.5 cm}\non\\
=\ {\rm e}^{\hs{0.03 cm}2\pi^{-1}\hs{-0.06 cm}\int(\p_z{\phi})
(\bar{\p}_{\bar z}{\phi})
\hs{0.05 cm}d^2z\hs{0.03 cm}
+\hs{0.03 cm}(2\pi i)^{-1}\hs{-0.08 cm}\int{\phi}{\cal R}}
\hs{0.2 cm}{\rm det}\hs{-0.04 cm}
\left((\bar\p+\frac{_1}{^2}\tau_2^{-1}Ud\bar z)^{^*}\hs{0.04 cm}
(\bar\p+\frac{_1}{^2}\tau_2^{-1}Ud\bar z)\right)^{-1}\
\en
\no where $\cal R$ denotes the metric curvature form.
Rather surprisingly, the resulting effective $\phi$
theory is the free field with the background charge so
that we obtain again a calculable functional integral.
Eq.\hs{0.05 cm}(8) follows by the standard chiral anomaly
calculation and does not depend on the regularization
scheme used to define the determinants,
within a large class. In particular, the absence of
the Liouville $\sim\hs{-0.04 cm}
\int\hs{-0.05 cm}{\rm e}^{\phi}$ term in the effective action,
of the type appearing in the conformal
anomaly calculation, is here not an artifact of the choice
of the zeta function regularization.
\vs 0.2 cm

The presence of the (generic) twist $U$ breaks the global $SL(2,\bC)$
symmetry of the theory to the diagonal $U(1)^\bC$.
The remaining symmetry results, however, in the divergence
of the $\phi$-integral (and, consequently, of the
partition function) due to the zero mode contribution.
This divergence may be extracted in the usual
way as the infinite volume
of $U(1)^\bC$ leading to the insertion of $\delta(\phi(0))$
fixing the $\phi$ zero mode under the integral.
The total central charge of the theory is easily computable
from the standard dependence of the resulting
determinants on the conformal factor of the metric. It is equal
$\hs{0.04 cm}c_{-\kappa}\equiv 3\kappa/(\kappa-2)\hs{0.04 cm}$
or\hs{0.14 cm}$\kappa\hs{0.05 cm}{\rm dim}\hs{0.04 cm}H/
(\kappa-h^\vee)\hs{0.04 cm}$ for general $H$. The determinants
are well known \cite{raysing}.
The final result is (in the flat metric; $q\equiv{\rm e}^{\hs{0.03 cm}
2\pi i\tau}\hs{0.04 cm}$):
\be
{\cal Z}^{H^+_3}(\tau,U)\ =\ C
\tau_2^{-1/2}\hs{0.05 cm}q\bar q\hs{0.03 cm}^{-1/8}\hs{0.06 cm}
\exp\left[{\hs{0.04 cm}-\pi(\kappa-2)U_2^2/\tau_2}\right]
|\sin(\pi U)|^{-2}\ \ \hs{-0.9 cm}\ \non\\
\cdot\hs{0.07 cm}
\bigg|\prod\limits_{n=1}^{\infty}(1-{\rm e}^{2\pi i\hs{0.04 cm}U}
\hs{-0.05 cm}q^n)
(1-q^n)(1-{\rm e}^{-2\pi i\hs{0.04 cm}U}\hs{-0.05 cm}q^n)
\bigg|^{-2}\ .\hs{-0.5 cm}
\en
\vs 0.2 cm

\no\underline{3.2.\ \ Quantum-mechanical model}
\vs 0.2 cm

It will be useful to interpret  expression (9) in the hamiltonian
language. Let us first do it in the simpler quantum-mechanical case
obtained from field theory by taking field configurations
independent of the space coordinate (this approximation, ignoring
the contributions of stringy oscillations, has been widely used
in 2D gravity where it goes under the catchy name of ``mini
superspace''). The quantum-mechanical system that we obtain here
is the geodesic motion on $H_3^+$ with the euclidean action
\be
-S_{\rm mini}(hh^\D)=\frac{_\kappa}{^4}\int{\rm tr}\hs{0.05 cm}
((hh^\D)^{-1}\p_t(hh^\D))^2\hs{0.05 cm}dt\ .
\en
\no Unlike in the 2D theory, in the ``mini'' case also
the real time action
is real and unitarity is recovered. The space of states is
$L^2(H_3^+,d(hh^\D))\cong L^2({\bf R}\times\bC,d\phi
\hs{0.04 cm}d^2v)$ and it carries the unitary representation
of $SL(2,\bC)$ defined by
\be
(gf)(hh^\D)=f(g^{-1}hh^\D{g^\D}^{-1})\ .
\en
\no On the infinitesimal level, this action may be described
by generators of $sl(2,\bC)\oplus sl(2,\bC)$
$\cong sl(2,\bC)^{\bC}$\hs{0.13 cm}:
\vs 0.4 cm

\vbox to 1.7 cm{
\hsize = 16.0 cm
\hs{1.4 cm}$J^1=\frac{_1}{^4}(1+|v|^2)^{-1/2}(v{\rm e}^{\phi}
-\bar v{\rm e}^{\phi})\p_\phi
-\frac{_1}{^2}(1+|v|^2)^{1/2}({\rm e}^\phi\p_{\bar v}+
{\rm e}^{-\phi}\p_v)\ ,$

\hs{1.4 cm}$J^2=\frac{_i}{^4}(1+|v|^2)^{-1/2}(v{\rm e}^{\phi}
+\bar v{\rm e}^{\phi})\p_\phi
-\frac{_i}{^2}(1+|v|^2)^{1/2}({\rm e}^\phi\p_{\bar v}-
{\rm e}^{-\phi}\p_v)\ ,$

\hs{1.4 cm}$J^3=-\frac{_1}{^2}\p_\phi-\frac{_1}{^2}v\p_v
+\frac{_1}{^2}\bar v\p_{\bar v}\ ,$
\vfill}
\no satisfying $[J^a,J^b]=i\epsilon^{abc}J^c$ and by $\bar J^a$'s
given by the complex-conjugate vector fields.
${J^a}^*=-\bar J^a$ so that
$J^a-\bar J^a$ and $i(J^a+\bar J^a)$ are the hermitian
generators of $sl(2,\bC)$. The Hamiltonian may be taken as
$-2\kappa^{-1}\Delta$ where $\Delta$ denotes the Laplace-Beltrami
operator on $H^+_3$ with the $SL(2,\bC)$-invariant metric.,
\be
 \Delta={\vec{J}}^{2\atop{}}={\vec{\bar J}}^{_2}=\frac{_1}{^{4}}\p_\phi^2
-\frac{_1}{^{4}}|v|^2(1+|v|^2)^{-1}\p_\phi^2
\ \ \ \ \ \ \ \non\\
+\hs{0.06 cm}(1+|v|^2)\p_v\p_{\bar v}
+\frac{_1}{^{4}}(v\p_v-\bar v\p_{\bar v})^2
+\frac{_1}{^{2}}(v\p_v
+\bar v\p_{\bar v})\ .
\en
\no $-\Delta$ has continuous bounded below
spectrum starting from $\frac{_1}{^4}$
and induces the decomposition
\be
L^2(H^+_3)\cong\int\limits_{\rho>0}\hs{-0.18 cm}^{^{^{^\bigoplus}}}
\hs{0.09 cm}{\cal H}_\rho\hs{0.09 cm}\rho^2d\rho
\en
\no into the direct integral of irreducible unitary
representations of $SL(2,\bC)$ from the principal continuous series
\cite{gelf5},\cite{vilen}
on which $-\Delta$ acts as multiplication by $(1+\rho^2)/4$.
${\cal H}_\rho$ may be realized as the space of homogeneous functions
of degree $-1+i\rho$ on non-negative matrices
$h'{h'}^\D$ with determinant zero, i.e.
on the upper light cone $V^+_3$. The parametrization by
$(\phi,v)$ together with all the formulae concerning the
action of $SL(2,\bC)$ pass to the case of $V^+_3$ provided
that we replace everywhere\footnote{this is
why formula (12) was written in a clumsy way}
factor $1+|v|^2$ by $|v|^2$.
The scalar product in ${\cal H}_\rho$ is that
of $L^2(\hs{0.06 cm}\delta(2-\hs{0.05 cm}
{\rm tr}\hs{0.05 cm}h'{h'}^\D\hs{0.06 cm})\hs{0.06 cm}
d(h'{h'}^\D)\hs{0.06 cm})$. Operators
$J^3-\bar J^3=-v\p_v+\bar v\p_{\bar v}$ and $i(J^3+\bar J^3)=-i\p_\phi$
may be diagonalized
at the same time as $\Delta$ and their joint spectrum is
${\bf Z}\times{\bf R}$ in each ${\cal H}_\rho$ which, consequently,
is very different from the highest- or lowest-weight
representation spaces of $sl(2,\bC)\oplus sl(2,\bC)$:
both $J^3$ and $\bar J^3$ have continuous unbounded spectrum
here!
\vs 0.2 cm

The heat kernel on $H^+_3$ is known explicitly and it has a
simple form:
\be
{\rm e}^{t\Delta}(h_1h_1^\D,h_2h_2^\D)=
(\pi t)^{-3/2}\frac{_d}{^{{\rm sinh}\hs{0.03 cm}d}}
\hs{0.05 cm}{\rm e}^{-t/4-d^2/t}
\en
\no where $d$ is the hyperbolic distance between $h_1h_1^\D$
and $h_2h_2^\D$ or between $hh^\D$ and $1$ where $h=h_2^{-1}h_1$.
In the more standard parametrization of $H^+_3$
\be
hh^\D=(1+{\vec{x}}^2)^{1/2}+\vec{x}\cdot\vec{\sigma}\ ,
\en
\no $d={\rm sinh}^{-1}(|\vec x|)$. Operator
${\rm e}^{t\Delta}$ is certainly not of trace class since
$-\Delta$ has continuous spectrum and moreover
of infinite multiplicity. In the formal expression
\be
\int{\rm e}^{t
\Delta}({\rm e}^{-\pi iU\sigma^3}hh^\D
{\rm e}^{\pi i\bar U\sigma^3}\hs{0.05 cm},\hs{0.05 cm}hh^\D)
\hs{0.08 cm}d(hh^\D)
\en
\no for $\hs{0.05 cm}{\rm tr\hs{0.07 cm}e}^{t\Delta}
\hs{0.06 cm}{\rm e}^{2\pi i(UJ^3-\bar U\bar J^3)}$, the integral
diverges due to the $U(1)^\bC$ symmetry of the integrated
kernel. That is the familiar problem which we have encountered
already in the two-dimensional theory. We
solve it again by fixing the $U(1)^\bC$ invariance in the
standard fashion. This leads to the insertion of
$\delta(\phi)$ under the integral of the right hand side
of (16)
which renders it finite (for $U\not\in
{\bf Z}$). The hyperbolic distance between
${\rm e}^{-\pi iU\sigma^3}hh^\D
{\rm e}^{\pi i\bar U\sigma^3}$ and $hh^\D$
\be
d=\cosh^{-1}\left((1+|v|^2)\cosh(2\pi U_2)
-|v|^2\cos(2\pi U_1)\right)
\en
\no for $hh^\D=\left(\matrix{(1+|v|^2)^{1/2}&v\cr
\bar v&(1+|v|^2)^{1/2}\cr}\right)$ and an easy calculation gives
\be
{\rm tr_{\rm ren}\hs{0.07 cm}e}^{4\pi \tau_2\kappa^{-1}\Delta}
\hs{0.06 cm}{\rm e}^{2\pi i(UJ^3-\bar U\bar J^3)}\ \equiv\
\int{\rm e}^{4\pi\tau_2\kappa^{-1}
\Delta}({\rm e}^{-\pi iU\sigma^3}hh^\D
{\rm e}^{\pi i\bar U\sigma^3}\hs{0.05 cm},\hs{0.05 cm}hh^\D)
\hs{0.08 cm}\delta(\phi)\hs{0.08 cm}d(hh^\D)\ \non\\
=\ \frac{_{\kappa^{1/2}}}{^{8\pi\tau_2^{1/2}}}{\rm e}^{-\pi
\tau_2/\kappa-\pi\kappa U_2^2/\tau_2}\hs{0.04 cm}
|\sin(\pi U)|^{-2}\ .
\en
\no On the other hand, the quantum-mechanical partition
function
\be
{\cal Z}_{\rm mini}^{H^+_3}
(\tau,U)\ =\ \int{\rm e}^{\kappa S_{\rm mini}
(hh^\D)}\hs{0.08 cm}\delta(\phi(t_0))\hs{0.08 cm}D(hh^\D)\non
\en
\no over twisted paths on $[0,2\pi\tau_2]$ satisfying
$hh^\D(2\pi\tau_2)={\rm e}^{-\pi iU\sigma^3}hh^\D(0)
{\rm e}^{\pi i\bar U\sigma^3}$
may be again computed by iterative gaussian integration.
Not too surprisingly, one finds
\be
{\cal Z}_{\rm mini}^{H^+_3}(\tau,U)\ =\ C
\tau_2^{-1/2}\hs{0.05 cm}{\rm e}^{-\pi\kappa U_2^2/\tau_2}
\hs{0.08 cm}|\sin(\pi U)|^{-2}\ .
\en
Comparing eqs.\hs{0.05 cm}(18) and (19), we find that
\be
{\cal Z}_{\rm mini}^{H^+_3}(\tau,U)\ =\ C\hs{0.06 cm}
{\rm tr_{\rm ren}\hs{0.07 cm}e}^{4\pi \tau_2\kappa^{-1}
(\Delta+1/4)}\hs{0.06 cm}{\rm e}^{2\pi
i(UJ^3-\bar U\bar J^3)}
\en
\no which establishes a Feynman-Kac type formula for the
hyperbolic space $H_3^+$. Similar formulae may be produced
for other symmetric spaces $H^\bC/H$.
\vs 0.3 cm

\no\underline{3.3.\ \ Space of states}
\vs 0.2 cm

Let us return now to the interpretation of expression
(9) for the 2D partition function which becomes now
straightforward. Using eq. (19) and (20), we obtain
\be
{\cal Z}^{H^+_3}(\tau,U)\ =\ C\hs{0.06 cm}
q\bar q\hs{0.03 cm}^{-c_{-\kappa}/24}\hs{0.1 cm}
{\rm tr_{\rm ren}\hs{0.07 cm}e}^{4\pi \tau_2(\kappa-2)^{-1}
\Delta}\hs{0.06 cm}{\rm e}^{2\pi
i(UJ^3-\bar U\bar J^3)}\ \ \non\\
\cdot\hs{0.07 cm}
\bigg|\prod\limits_{n=1}^{\infty}(1-
{\rm e}^{2\pi iU}\hs{-0.04 cm}q^n)
(1-q^n)(1-{\rm e}^{-2\pi iU}\hs{-0.04 cm}q^n)
\bigg|^{-2}\ .
\en
\no The first term on the right is the familiar
prefactor with the central charge. Next comes essentially
the mini-space contribution with
$\kappa\mapsto\kappa-2$ and then, multiplicatively,
the contribution of the oscillatory degrees of freedom.
By studying the canonical quantization of the $H^+_3$
WZW theory, one may infer that its space of states should
carry a representation of the affine algebra
$\hat{sl}(2,\bC)\oplus\hat{sl}(2,\bC)$ of level $-\kappa$,
extending
the mini-space representation of $sl(2,\bC)\oplus
sl(2,\bC)$. Let $\hat{b}_{\pm}$ ($\hat{n}_{\pm}$)
denote the subalgebras of $\hat{sl}(2,\bC)$ generated by
$J^a_n$ with $\pm n\geq 0$ ($\pm n>0$). The
action of $sl(2,\bC)\oplus sl(2,\bC)$
in $L^2(H^+_3)$  may be extended to a representation
of $\hat{b}_+\oplus\hat{b}_+$ by making $J^a_n$ and
$\bar J^a_n$ for $n>0$ act trivially (the bar refers
to the second copy). Let us choose a dense invariant
subdomain in $L^2(H^+_3)$ like the space ${\cal S}
(H^+_3)$ of fast decreasing functions (in $\vec x$ of (15)).
$\hat{sl}(2,\bC)\oplus\hat{sl}(2,\bC)$
acts then in the space
\be
\hat{\cal H}^{H^+_3}\ =\ \bigg({\cal U}(\hat{sl}(2,\bC))\otimes
{\cal U}(\hat{sl}(2,\bC))\bigg)\otimes_{{\cal U}(\hat{b}_+)
\otimes{\cal U}(\hat{b}_+)}{\cal S}(H^+_3)
\en
\no where $\cal U$ denotes the enveloping algebra.
This gives the representation of
$\hat{sl}(2,\bC)\oplus\hat{sl}(2,\bC)$ induced
from the action of $sl(2,\bC)\oplus sl(2,\bC)$
in $L^2(H^+_3)$. In plain English, space $\hat{\cal H}^{H^+_3}$
is spanned by ${\cal S}(H^+_3)$ and by the descendents
obtained by repeated action of $J^a_n$ and $\bar J^b_n$ with
$n<0$ on the states in ${\cal S}(H^+_3)$. As a vector space,
\be
\hat{\cal H}^{H^+_3}\cong {\rm Sym}(\hat{n}_-)
\otimes{\rm Sym}(\hat{n}_-)\otimes{\cal S}(H^+_3)\non
\en
\vs 0.01 cm
\no where \hs{0.06 cm}Sym\hs{0.06 cm}
denotes the symmetric algebra.
As usually,
the Sugawara construction allows to define the action
in $\hat{\cal H}^{H^+_3}$ of two commuting Virasoro algebras
(of central charge $c_{-\kappa}$):
\vs 0.01 cm
\be
L_n=-\frac{_1}{^{\kappa-2}}\sum\limits_{m,a}
:J^a_mJ^a_{n-m}:
\en
\no and similarly for $\bar L_n$. It is then the standard
result that the contribution of the descendent states to
$\hs{0.09 cm}{\rm tr}\hs{0.07 cm}q^{L_0}\bar q^{\bar L_0}
{\rm e}^{4\pi i(UJ^3_0-\bar U\bar J^3_0)}\hs{0.09 cm}$
is the infinite product factor in (21). Since
\be
q^{L_0}\bar q^{\bar L_0}\hs{0.07 cm}\bigg|_{L^2(H^+_3)}=
{\rm e}^{4\pi\tau_2(\kappa-2)^{-1}\Delta}\ ,
\en
\vs 0.01 cm
\no also the (renormalized) zero-level states
contribution is recovered in (21).
\vs 0.24 cm

The hamiltonian interpretation of the field-theoretic
partition function may be then summarized in the following
(Feynman-Kac type) formula:
\be
{\cal Z}^{H^+_3}(\tau,U)\ =\ q\bar q^{\hs{0.03 cm}-c_{-\kappa}/24}
\hs{0.08 cm}{\rm tr_{ren}}\hs{0.07 cm}
q^{L_0}\bar q^{\bar L_0}
{\rm e}^{2\pi i(UJ^3_0-\bar U\bar J^3_0)}
\en
\no where on the right hand side the (renormalized) trace
is taken over the space $\hat{\cal H}^{H^+_3}$ carrying the
representation of $\hat{sl}(2,\bC)\oplus\hat{sl}(2,\bC)$
induced from $L^2(H^+_3)$. The structure of the partition
function of (21) and of the space of states appears
to be much simpler here than in the case of compact WZW
models. The probable reason is that
$\hat{\cal H}^{H^+_3}$ may be decomposed into a direct integral of
representations induced from ${\cal H}_\rho$, which we expect to be
irreducible, at least in a suitable sense and for almost
all $\rho$. Similar decomposition in the compact case
(into a finite direct sum)
yields representations which should be further reduced.
$\hat{\cal H}^{H^+_3}$ carries a natural
hermitian form $(\ \hs{0.07 cm},\ )$ extending the scalar product of
$L^2(H^+_3)$.
It may be characterized by the conjugacy relation
${J^a_n}^{*}=-\bar J^a_{-n}$. It is certainly non-positive
since for $\chi\in L^2(H^+_3)$
\be
(\hs{0.05 cm}(J^1_{-1}-\bar J^1_{-1})\chi)\hs{0.05 cm},
\hs{0.05 cm}(J^1_{-1}-\bar J^1_{-1})\chi\hs{0.05 cm})
=-\frac{_\kappa}{^2}(\hs{0.05 cm}\chi\hs{0.05 cm},
\hs{0.05 cm}\chi\hs{0.05 cm})\ .
\en
\no We expect however that $(\ \hs{0.07 cm},\ )$ is non-degenerate.
\vs 0.3 cm

\no\underline{3.4.\ \ Green functions}
\vs 0.2 cm

In Sec. 2.1 and 2.2, we have seen that the matrix elements
$hh^\D(\xi)_j$ of spin $j=0,\frac{_1}{^2},1,...$
representations appear as
natural insertions in the $SL(2,\bC)/SU(2)$ WZW theory,
provided that they are arranged into combinations invariant
under the global $SL(2,\bC)$ symmetry
$hh^\D\mapsto\gamma hh^\D\gamma^\D$ (this is like the
neutrality condition in the 2D Coulomb gas correlations).
The corresponding Green functions are calculable by the iterative
gaussian integration in parametrization (1). Let us explain
how this works on the simplest example of the planar
spin $\frac{_1}{^2}$ two-point function \cite{quad}
\be
\int\left({\rm tr}_{1/2}\hs{0.06 cm}hh^\D(\xi_1)\hs{0.06 cm}
(hh^\D)^{-1}(\xi_2)\right)\hs{0.1 cm}{\rm e}^{\hs{0.03 cm}\kappa
\hs{-0.05 cm}\int\hs{-0.05 cm}S(hh^\D)}
\hs{0.07 cm}\delta(hh^\D(\xi_0))\hs{0.07 cm}
D(hh^\D)\hs{0 cm}\non\\
=\ \int(|({\rm e}^{\phi}v)(\xi_1)-({\rm e}^{\phi}v)(\xi_2)|^2+
{\rm e}^{\phi(\xi_1)-\phi(\xi_2)}
+{\rm e}^{\phi(\xi_2)-\phi(\xi_1)})\hs{0 cm}\non\\
\cdot\hs{0.11 cm}{\rm e}^{\hs{0.03 cm}-
\pi^{-1}\kappa\int[\hs{0.05 cm}
(\p_z\phi)(\p_{\bar z}\phi)+
(\p_z+\p_z\phi)\bar v\hs{0.08 cm}
(\p_{\bar z}+\p_{\bar z}\phi)v)\hs{0.05 cm}]\hs{0.07 cm}
d^2z}\hs{0 cm}\non\\
\cdot\hs{0.11 cm}\delta(\phi(\xi_0))\hs{0.09 cm}
\delta^2(v(\xi_0))\hs{0.09 cm}D\phi\hs{0.08 cm}Dv
\en
\no where we have already shifted $\tilde\phi\mapsto\phi$.
The $v$-integral is gaussian.
It produces the partition function
\be
{\rm e}^{\hs{0.03 cm}2\pi^{-1}\hs{-0.08 cm}\int(\p_z{\phi})
(\bar{\p}_{\bar z}{\phi})
\hs{0.05 cm}d^2z\hs{0.05 cm}
+(2\pi i)^{-1}\hs{-0.05 cm}\int\hs{-0.05 cm}{\phi}{\cal R}}\hs{0.08 cm}
\left({\rm det}'(\bar\p^*\bar\p)/{\rm area}\right)^{-1}
\en
\no (which changes the coupling
constant of the effective $\phi$-integral from $\kappa$ to
$\kappa-2$, compare eq. (8)) and the normalized expectation
\be
<|({\rm e}^{\phi}v)(\xi_1)-({\rm e}^{\phi})(\xi_2)|^2>\hs{5.7 cm}\non\\
=\ (\pi\kappa)^{-1}|\xi_1-\xi_2|^2\hs{0.12 cm}
{\rm e}^{-\phi(\xi_1)-
\phi(\xi_2)}\int{\rm e}^{2\phi(\zeta)}
|\xi_1-\zeta|^{-2}|\xi_2-\zeta|^{-2}\hs{0.05 cm}d^2\zeta\ .
\en
\no Notice the appearance of the linear term $\sim\smallint\phi
{\cal R}$ in the effective
$\phi$-action and of the ${\rm e}^{2\phi(\zeta)}$
insertion corresponding, respectively, to the background
and screening charges in the Coulomb gas interpretation
of the resulting $\phi$-field theory. The integral over $\phi$
is again gaussian but requires a
renormalization of the polynomial in ${\rm e}^{\pm\phi(\xi_\alpha)}$
and ${\rm e}^{2\phi(\zeta)}$ to render it finite. If we extract
the most divergent factor multiplicatively, the terms with
milder divergences will not survive the renormalization. In
the case at hand, these are terms ${\rm e}^{\phi(\xi_1)-\phi(\xi_2)}
+{\rm e}^{\phi(\xi_2)-\phi(\xi_1)}$ on the right hand side of (27).
They drop out leaving us with the result
\be
{\rm const.}\hs{0.09 cm}|\xi_1-\xi_2|^{2-1/(\kappa-2)}\int|(\xi_1-\zeta)
(\xi_2-\zeta)|^{-2+2/(\kappa-2)}\hs{0.05 cm}d^2\zeta\non\\
=\ {\rm const.}\hs{0.09 cm}|\xi_1-\xi_2|^{3/(\kappa-2)}
\en
\no (in the flat metric). Replacing ${\rm tr}_{\frac{_1}{^2}}$ in
(27) by ${\rm tr}_{j}$
for higher spins, we obtain a $\phi$-integral with
$2j$ screening charges and finally
\be
{\rm const.}\hs{0.09 cm}|\xi_1-\xi_2|^{4j\hs{0.03 cm}(j+1)/(\kappa-2)}
\en
\no provided that $2j+1<\kappa-2$. Otherwise, the integral
over the positions of the screening charges diverges\footnote{
this is the dual manifestation of the restriction to spins $j\leq k/2$
in the $SU(2)$ WZW model or, more generally, of its fusion rules,
see \cite{quad}}. Higher Green functions may be computed similarly
\cite{quad},\cite{geom}, also for the general $H^\bC/H$ theories \cite{fgk}.
\vs 0.2 cm

{}From the form of the general Green functions (also with the
current and energy-momentum insertions) one infers that
fields $hh^\D(\xi)$ are primary, both for the $\hat{sl}(2,\bC)
\oplus\hat{sl}(2,\bC)$ and Vir$\oplus$Vir algebras. Their
conformal weights $\Delta_j=\bar\Delta_j$
are, as read from eq. (31), $-\frac{{j(j+1)}}{{\kappa-2}}<0$.
Occurrence of fields with negative dimensions, so with
Green functions growing with the distance,
might seem incompatible
with the stability although not necessarily in a
non-unitary theory as ours. The point, however, lies
elsewhere. Such fields (:${\rm e}^{\alpha\phi}$: for $\alpha$ real)
are clearly present for the massless free (uncompactified) field
$\phi$ which gives a stable unitary theory and are also expected
in the Liouville theory \cite{seib}, believed to be stable
and unitary (there, they correspond to the local operators in
terminology of \cite{seib}). These operators escape the
standard relation between the spectrum of energy and of
conformal weights since they correspond to eigenfunctions
of the Hamiltonian outside the generalized eigenspaces.
This may be seen already in the ``mini-space'' quantum-mechanical
picture which is stable and unitary for the $H^+_3$ theory:
although
\be
-\frac{_1}{^{\kappa-2}}\Delta\hs{0.06 cm}hh^\D_j\ =\
-\frac{_{j(j+1)}}{^{\kappa-2}}\hs{0.05 cm}hh^\D_j\ ,\non
\en
\no the matrix elements of $hh^\D_j$ are not the generalized
eigenfunctions of $-\Delta$ due to their too rapid growth
at infinity. Appearance of operators with negative conformal
dimensions may be typical for irrational theories with
continuous spectrum of $L_0,\bar L_0$. Notice nevertheless
that in the $H^\bC/H$ WZW model they come in a finite number
whereas for the massless free field and for the Liouville theory,
there is a continuous family of such fields.
\vs 0.2 cm

Besides fields with negative dimensions which do not correspond
neither to true nor to generalized states of the theory, it
is natural to expect existence of fields with positive
dimensions corresponding to the states in the spectrum
of $L_0,\bar L_0$. The natural candidates for such fields
are given by $f_{\rho,m_l,m_r}(hh^\D(\xi))$ where $f_{\rho,m_l,m_r}$
is a joint generalized eigenfunction of $-\Delta,J^3,\bar J^3$
corresponding to eigenvalues $\frac{_1}{^4}(1+\rho^2),
m_l=\frac{_1}{^2}(n+i\omega), m_r=\frac{_1}{^2}(-n+i\omega)$
where $\rho\geq 0$, $n\in{\bf Z}$ and $\omega
\in{\bf R}$. In the space ${\cal H}_\rho$ (of homogenous functions on
$V^+_3$), the corresponding eigenfunction is
\be
{\rm e}^{-i\omega\phi-in\hs{0.03 cm}{\rm arg}\hs{0.03 cm}(v)}
\hs{0.06 cm}|v|^{-1+i\rho}\ .
\en
\no Eigenfunction $f_{\rho,m_l,m_r}$ on $H^+_3$ is
obtained by applying to (32) the
Gelfand-Graev integral transformation \cite{gelf5} realizing
the isomorphism (13):
\be
f_{\rho,m_l,m_r}(\phi,v)
\ =\ {\rm e}^{-i\omega\phi-in\hs{0.03 cm}{\rm arg}\hs{0.03 cm}(v)}
\hs{0.06 cm}(1+|v|^2)^{i\omega/2}\hs{1.5 cm}\non\\
\cdot\hs{0.1 cm}\int\limits_0^{2\pi}d\theta
\int\limits_0^\infty dr\hs{0.13 cm}{\rm e}^{in\theta}
\hs{0.06 cm}r^{i\rho+i\omega}\hs{0.06 cm}[\hs{0.04 cm}
1+2|v|r{\rm cos}\theta\hs{0.04 cm}
+(1+|v|^2)r^2\hs{0.04 cm}]^{-1-i\rho}\ .
\en
\no For example, for $\rho=m_l=m_r=0$, we obtain
the elliptic integral
\be
f_{0,0,0}(v)\ =\ \pi\int\limits_0^{2\pi}
(1+|v|^2{\rm sin}^2\theta)^{-1/2}\hs{0.07 cm}d\theta\ .
\en
\no Unfortunately, we were not able to compute the Green functions
of fields $f_{\rho,m_l,m_r}$ exactly. It remains then
to be seen if they indeed give rise, upon multiplicative
renormalization, to primary fields with conformal
weights $\Delta_\rho=\bar\Delta_\rho=
\frac{{1+\rho^2}}{{4(\kappa-2)}}$.
\vs 0.5 cm

\no 4.\ \ $SL(2,\bC)/SU(2)$ mod $\bf R$ COSET THEORY
\vs 0.3 cm
\no\underline{4.1\ \ 2D black hole sigma model}
\vs 0.2 cm
\addtocounter{equation}{-34}

In Sec. 3, we have coupled the $SL(2,\bC)/SU(2)\equiv H^+_3$
WZW model to an abelian gauge field $A$ in the way which rendered
the action invariant under the non-compact gauge transformations:
\be
S({\rm e}^{\lambda\sigma^3/2}hh^\D{\rm e}^{\lambda\sigma^3/2},
\frac{_1}{^{2i}}(A-d\lambda))\ =\ S(hh^\D,\frac{_1}{^{2i}}A)\ ,
\en
\no see (3.3). Following the scenario for producing coset
theories from compact WZW models, let us consider the functional
integral
\be
\int\ -\ \ {\rm e}^{\hs{0.05 cm}\kappa S(hh^\D,\hs{0.04 cm}
(2i)^{-1}A)}\hs{0.06 cm}
D(hh^\D)\hs{0.06 cm}DA\hs{6 cm}\non\\
=\ \int\ -\ \ {\rm e}^{\hs{0.04 cm}-\pi^{-1}\kappa\int[(\p_z\tilde\phi
+A_z)(\p_{\bar z}\tilde\phi+A_{\bar z})+(\p_z+\p_z\tilde\phi
+A_z)\bar v\hs{0.04 cm}(\p_{\bar z}+\p_{\bar z}\tilde\phi+A_{\bar z})v]
\hs{0.04 cm}d^2z}\hs{0.07 cm}D\phi\hs{0.05 cm}
Dv\hs{0.05 cm}DA
\en
\no with gauge invariant insertions. First notice that, by the gauge
invariance, the integral over $\phi$ factors as the (infinite)
volume of the gauge group. Since $A$ enters quadratically into
the action, it may be integrated out (for appropriate, e.g.
$A$-independent, insertions) giving
\be
C\int\ -\ \ {\rm e}^{\hs{0.03 cm}-\pi^{-1}\kappa\int
(1+|v|^2)^{-1}(\p_z\bar v)(\p_{\bar z}v)\hs{0.04 cm}d^2z}
\hs{0.07 cm}\prod\limits_\xi\frac{_{d^2v(\xi)}}{^{1+|v(\xi)|^2}}\ .
\en
\no The effective action for $v$:
\be
S_{\rm eff}(v)\ \equiv\ \frac{_{\kappa}}{^{\pi}}\int
(1+|v|^2)^{-1}(\p_z\bar v)(\p_{\bar z}v)\hs{0.05 cm}
d^2z\non\\
=\ \frac{_\kappa}{^\pi}\sum\limits_{a=1,2}\int
(1+|v|^2)^{-1}(\p_zv^a)(\p_{\bar z}v^a)\non
\hs{0.05 cm}d^2z
\en
\no if we integrate by parts. $v=v^1+iv^2$. It is the action of
a sigma model with the complex plane with metric
\be
(1+|v|^2)^{-1}(dz\otimes d\bar z+
d\bar z\otimes dz)
\en
\no as the target. It was noticed recently
\cite{w3},\cite{msw} that this target metric
(together with the dilaton field $\Phi={\rm log}(1+|v|^2)$\hs{0.04 cm})
forms a euclidean
black hole solution of equations of 2D gravity (with unit mass).
It describes an infinite cigar becoming asymptotically a
cylinder (the scalar curvature goes down as $|v|^{-2}$
as $v\rightarrow\infty$). The Minkowskian counterpart of this
solution is the metric
\be
(1-v^+v^-)^{-1}(dv^+dv^-+dv^-dv^+)
\en
\no with the asymptotically flat region $\pm v^{\pm}>0$
with future horizon $v^-=0,\ v^+>0$ and past horizon
$v^+=0,\ v^-<0$, another such region for $v^+\leftrightarrow v^-$,
and future and past singularities at $v^+v^-=1$.
\vs 0.3 cm

\no\underline{4.2\ \ Toroidal partition function}
\vs 0.2 cm

As it stands, functional integral (3) for the black
hole target is difficult to
compute directly. Instead, we may go back to expression (2) and
integrate first over $hh^\D$ and then over $A$. Let us illustrate
this on the example of the twisted toroidal partition function
\be
{\cal Z}^{\rm bh}(\tau,U)\ =\ \int{\rm e}^{\hs{0.05 cm}\kappa
S(\gamma_Uhh^\D\gamma_U^\D,\hs{0.08 cm}
(2i)^{-1}\hs{-0.05 cm}A)}\hs{0.06 cm}
D(hh^\D)\hs{0.06 cm}DA\hs{3 cm}
\en
\no where the action for the twisted field configurations is coupled
to the gauge field by putting
\be
S(\gamma_Uhh^\dagger\gamma_U^\dagger,\frac{_1}{^{2i}}A)\ =
\ S(hh^\dagger\hs{0.05 cm},\hs{0.04 cm}\frac{_1}{^{2i}}(A+
\frac{_1}{^2}\tau_2^{-1}\bar Udz+\frac{_1}{^2}
\tau_2^{-1}Ud\bar z))\ \ \non\\
+\ \frac{_1}{^{2\pi\tau_2}}U_1\int(A_z+
A_{\bar z})d^2z\ +\ \frac{_{\pi}}{^{\tau_2}}U_1^2\ .
\en
The parametrization of $A$ by the Hodge decomposition
\be
A=d\mu+*d\nu+\tau_2^{-1}(\bar udz+ud\bar z)/2
\en
\no ($\mu,\nu$ real functions, $u=u_1+iu_2$) gives
for the volumes
\be
DA\ =\ C\tau_2^{-2}\hs{0.04 cm}
{\rm det}'(\bar\p^*\bar\p)\hs{0.13 cm}\delta(\mu(\xi_0))
\hs{0.1 cm}\delta(\nu(\xi_0))\hs{0.12 cm}d^2u\hs{0.1 cm}
D\mu\hs{0.1 cm}D\nu\ .\non
\en
\no Due to the gauge invariance of the action, the integral
over $\mu$ factors out as the (infinite)
volume of the gauge group. The $\nu$-integral also factors out
after unitary rotation $v\mapsto {\rm e}^{\hs{0.02 cm}-i\nu}v$
so that the $v$- and $\phi$-integrals produce the
twisted partition function ${\cal Z}^{H^+_3}(\tau,u)$
of the $H^+_3$ WZW theory.
As the result, we obtain
\be
{\cal Z}^{\rm bh}(\tau,U)\ =\ C\tau_2^{-2}
\hs{0.05 cm}{\rm det}'(\bar\p^*\bar\p)
\int{\rm e}^{\hs{0.03 cm}
-\pi^{-1}\kappa\int(\p_z\nu)(\p_{\bar z}\nu)\hs{0.03 cm}d^2z
\hs{0.03 cm}-\hs{0.03 cm}\pi\kappa
\tau_2^{-1}(U_1-u_1)^2}\ \non\\
\cdot\hs{0.1 cm}
{\cal Z}^{H^+_3}(\tau,u)\hs{0.1 cm}d^2u\hs{0.1 cm}\delta(\nu(\xi_0))
\hs{0.07 cm}D\nu\ .
\en
\no The $\nu$-integral is straightforward and for
${\cal Z}^{H^+_3}(\tau,u)$ we have expression (3.9). Hence
\be
{\cal Z}^{\rm bh}(\tau,U)\ =\ C\tau_2^{-1/2}\int
{\rm e}^{\hs{0.03 cm}-\pi\kappa
\tau_2^{-1}(U_1-u_1)^2}\hs{0.05 cm}{\cal Z}
(\tau,u)\hs{0.08 cm}
|\eta(\tau)|^2\hs{0.09 cm}d^2u\non\\
=\ C\tau_2^{-1}\hs{0.05 cm}q\bar q\hs{0.07 cm}^{-1/12}\hs{0.02 cm}
\int{\rm e}^{\hs{0.05 cm}-\pi\kappa\tau_2^{-1}
(U_1-u_1)^2-\pi(\kappa-2)\tau_2^{-1}u_2^2}
\hs{0.1 cm}|\sin(\pi u)|^{-2}\non\\
\cdot\hs{0.1 cm}
\bigg|\prod\limits_{n=1}^{\infty}(1-{\rm e}^{2\pi iu}q^n)
(1-{\rm e}^{-2\pi iu}q^n)
\bigg|^{-2}\hs{0.1 cm}d^2u
\en
\no where $\eta(\tau)\equiv q^{1/24}\prod\limits_{n\geq 1}(1-q^n)$
is the Dedekind function. The $u$-integral
diverges logarithmically due to the singularity $\sim|u|^{-2}$
at zero. This singularity is repeated on the lattice ${\bf Z}+
\tau{\bf Z}$ \hs{0.01 cm}, \hs{0.01 cm} as follows
immediately from the bi-periodicity of expression\hs{0.1 cm}
${\rm e}^{2\pi\tau_2^{-1}u_2^2}\hs{0.06 cm}|{\rm sin}(\pi u)|^{-2}
\hs{0.06 cm}\bigg|\prod\limits_{n=1}^{\infty}(1-{\rm e}^{2\pi iu}q^n)
(1-{\rm e}^{-2\pi iu}q^n)\bigg|^{-2}$\hs{0.08 cm}.\hs{0.08 cm}
Let us explain this divergence of a relatively simple nature.
\vs 0.3 cm

\no\underline{4.3\ \ Mini-space partition function}
\vs 0.2 cm

It is instructive to start with the mini-space case (we remind that
this means taking fields $hh^\D$ and $A_z,
\hs{0.05 cm}A_{\bar z}$ independent of the space variable).
For
\be
{\cal Z}^{\rm bh}_{\rm mini}(\tau,U)\ =\ \int{\rm e}^{\hs{0.05 cm}\kappa
S_{\rm mini}(\gamma_Uhh^\D\gamma_U^\D\hs{0.05 cm},\hs{0.11 cm}
(2i)^{-1}\hs{-0.05 cm}A)}\hs{0.06 cm}
D(hh^\D)\hs{0.06 cm}DA\ ,
\en
\no we may also proceed as before integrating first over $A$
to get the twisted partition function for the quantum-mechanical
particle moving on the euclidean black hole:
\be
{\cal Z}^{\rm bh}_{\rm mini}(\tau,U)\ =\ C\int{\rm e}^{\hs{0.05 cm}
-(\kappa/2)\int\limits_0^{2\pi\tau_2}(1+|v|^2)^{-1}\hs{0.04 cm}|
(\p_t-i\tau_2^{-1}U_1)v|^2\hs{0.03 cm}dt}
\hs{0.07 cm}\prod\limits_\xi
\frac{_{d^2v(\xi)}}{^{1+|v(\xi)|^2}}\ .
\en
\no On the other hand, integrating first over $hh^\D$ and then
over $A$, we obtain:
\be
{\cal Z}^{\rm bh}_{\rm mini}(\tau,U)\ =\ C\tau_2^{-1}
\int{\rm e}^{\hs{0.05 cm}-\pi\kappa\tau_2^{-1}
((U_1-u_1)^2+u_2^2)}
\hs{0.1 cm}|\sin(\pi u)|^{-2}\hs{0.1 cm}d^2u\ .
\en
\no The right hand side of eq. (13) may be rewritten, with the use
of eqs. (3.18)-(3.20), as
\be
C\tau_2^{-1/2}\int{\rm tr_{ren}}\left({\rm e}^{\hs{0.05 cm}
4\pi\tau_2\kappa^{-1}(\Delta+1/4)}\hs{0.07 cm}{\rm e}^{\hs{0.04 cm}
2\pi i(uJ^3-\bar u\bar J^3)}\right)\hs{0.04 cm}{\rm e}^{\hs{0.05 cm}
-\pi\kappa\tau_2^{-1}(U_1-u_1)^2}\hs{0.14 cm}d^2u\hs{0.75 cm}\non\\
=C\tau_2^{-1/2}\int{\rm e}^{\hs{0.05 cm}4\pi\tau_2\kappa^{-1}
(\Delta+1/4)}\hs{0.04 cm}(2\pi u_2,{\rm e}^{-2\pi iu_1}v
\hs{0.05 cm};\hs{0.05 cm}0,v)\hs{0.15 cm}{\rm e}^{\hs{0.04 cm}
-\pi\kappa\tau_2^{-1}(U_1-u_1)^2}\hs{0.12 cm}d^2u\hs{0.07 cm}
d^2v\ .
\en
\no Notice that
\be
\int{\rm e}^{\hs{0.04 cm}t\Delta}\hs{0.02 cm}(2\pi u_2,v
\hs{0.05 cm};\hs{0.05 cm}0,v')\hs{0.09 cm}du_2
\ =\ \frac{_1}{^{2\pi}}\hs{0.05 cm}
{\rm e}^{\hs{0.05 cm}t\Delta_{\omega=0}}
\hs{0.04 cm}(v\hs{0.04 cm};\hs{0.04 cm}v')
\en
\no where $\Delta_{\omega=0}$ is the restriction of Laplacian
$\Delta$ to the generalized eigensubspace of operator $i(J^3+\bar J^3)
=-i\p_\phi$ corresponding to eigenvalue $0$. From the expression (3.12)
for $\Delta$, we infer that
\be
\Delta_{\omega=0}\ =\ (1+|v|^2)\p_v\p_{\bar v}+\frac{_1}{^4}
(v\p_v-\bar v\p_{\bar v})^2+\frac{_1}{^2}(v\p_v+\bar v
\bar\p_{\bar v})
\en
\no and is a selfadjoint operator in $L^2(d^2v)$.
Moreover,
\be
(\kappa/\tau_2)^{1/2}\int{\rm e}^{\hs{0.04 cm}4\pi\tau_2\kappa^{-1}
\Delta_{\omega=0}}({\rm e}^{\hs{0.02 cm}-2\pi i\hs{0.03 cm}u_1}v
\hs{0.05 cm};\hs{0.05 cm}v')\hs{0.13 cm}{\rm e}^{\hs{0.02 cm}
-\pi\kappa\tau_2^{-1}(U_1-u_1)^2}\hs{0.11 cm}du_1\hs{2.5 cm}\non\\
=\ (\kappa/\tau_2)^{1/2}\int{\rm e}^{\hs{0.04 cm}4\pi\tau_2\kappa^{-1}
\Delta_{\omega=0}\hs{0.04 cm}+\hs{0.04 cm}
2\pi i\hs{0.03 cm}u_1(J^3-\bar J^3)}(v
\hs{0.05 cm};\hs{0.05 cm}v')\hs{0.13 cm}{\rm e}^{\hs{0.02 cm}
-\pi\kappa\tau_2^{-1}(U_1-u_1)^2}\hs{0.11 cm}du_1\hs{1.8 cm}\non\\
=\ {\rm e}^{\hs{0.04 cm}4\pi\tau_2\kappa^{-1}
\Delta_{\omega=0}\hs{0.04 cm}-\hs{0.04 cm}
\pi\tau_2\kappa^{-1}(J^3-\bar J^3)^2
\hs{0.04 cm}+\hs{0.04 cm}2\pi i\hs{0.03 cm}U_1(J^3-\bar J^3)}\hs{0.04 cm}
(v\hs{0.05 cm};\hs{0.05 cm}v')\ =
\ {\rm e}^{\hs{0.05 cm}4\pi\tau_2\kappa^{-1}\Delta^{\rm bh}}
\hs{0.02 cm}({\rm e}^{\hs{0.03 cm}-2\pi i\hs{0.03 cm}U_1}
v\hs{0.05 cm};\hs{0.05 cm}v')\ \hs{0.1 cm}
\en
\no where we have introduced
\be
-\Delta_{\omega=0}+(J^3)^2\ =\ -\Delta_{\omega=0}+(\bar J^3)^2
\hs{1.6 cm}\non\\
=-\frac{_1}{^2}\p_v(1+|v|^2)\p_{\bar v}-\frac{_1}{^2}\p_{\bar v}
(1+|v|^2)\p_v\equiv-\Delta^{\rm bh}\ .
\en
\no It is a Laplacian quantizing the classical Hamiltonian
$p_vp_{\bar v}(1+|v|^2)$ of the particle on the (euclidean)
black hole, with a specific choice of ordering prescription
(different from the Laplace-Beltrami operator which would
correspond to $(1+|v|^2)^{1/2}\p_v\p_{\bar v}(1+|v|^2)^{1/2}
\hs{0.06 cm}).$ We may finally rewrite the mini-space
partition function as
\be
{\cal Z}^{\rm bh}_{\rm mini}(\tau,U)\ =\ C\int {\rm e}^{
\hs{0.05 cm}4\pi\tau_2\kappa^{-1}
(\Delta^{\rm bh}+1/4)}\hs{0.02 cm}({\rm e}^{\hs{0.03 cm}
-2\pi i\hs{0.03 cm}U_1}v\hs{0.05 cm};\hs{0.05 cm}v)\hs{0.14 cm}d^2v\ .
\en
\no The integral is divergent but the nature of this divergence
is quite simple. For $v\rightarrow\infty$, where the metric
becomes cylindrical in variable log$\hs{0.04 cm}
v$, $\exp[t\Delta^{\rm bh}
({\rm e}^{\hs{0.03 cm}-2\pi i\hs{0.03 cm}U_1}v\hs{0.04 cm};
\hs{0.04 cm}v)|v|^2$ approaches a constant (equal to
the free heat kernel between the points on the cylinder of
constant difference). Hence the divergence due to the
infinite volume of the black hole cigar. It may be easily
regularized by cutting integral over $v$ to $|v|\leq R$.
Going back to integral (3.8), it is easy to see that such cutoff
results in the replacement
\be
{\rm e}^{\hs{0.03 cm}-\pi\kappa\tau_2^{-1}u_2^2}
\ \longmapsto\ {\rm e}^{\hs{0.03 cm}-\pi\kappa\tau_2^{-1}u_2^2}-
{\rm e}^{\hs{0.03 cm}-(4\pi\tau_2)^{-1}\kappa\hs{0.03 cm}d_R^2}
\en
\no in the integrand of (13). Here $d_R=\cosh^{-1}
\left(\cosh(2\pi u_2)+2R^2|\sin(\pi u)|^2\right)$ stands for
the hyperbolic distance between ${\rm e}^{-\pi iU\sigma^3}hh^\D
{\rm e}^{\pi i\bar U\sigma^3}$ and $hh^\D
=\left(\matrix{(1+R^2)^{1/2}&R\cr R&(1+R^2)^{1/2}\cr}\right)$.
Such a replacement
makes the integral in (13) convergent but behaving as ${\cal O}
({\rm log}\hs{0.03 cm}R)$ (or more generally as ${\cal O}
({\rm log}\hs{0.03 cm}MR)$ where M is the black hole mass; we
consider here only the case $M=1$). We could define the finite part
of ${\cal Z}^{\rm bh}_{\rm mini}$ by subtracting this logarithmic
divergence, i.e. by comparing it to half the partition function
of a particle on the cylinder.
\vs 0.2 cm

Let us go back to the interpretation of the result (10).
As compared to expression (13) for the mini-space case,
the main differences in (10) are the partial shift
$\kappa\mapsto\kappa-2$ and the presence of the big
product inherited from the oscillatory modes of the
$H^+_3$ theory. The shift of $\kappa$ is easy:
if we drop the infinite product from the right hand side
of (10) to get the level zero (i.e. zero mode) contribution,
we obtain, proceeding as for the mini-space case,
\be
{\cal Z}^{\rm bh}_{{\rm level}\hs{0.08 cm}0}(\tau,U)\ =\
Cq\bar q^{\hs{0.04 cm}-(c_{-\kappa}-1)/24}\hs{0.18 cm}
{\rm tr}|_{\omega=0}
\hs{0.08 cm}{\rm e}^{\hs{0.03 cm}4\pi\tau_2(\kappa-2)^{-1}
\hs{-0.04 cm}\Delta\hs{0.06 cm}-\hs{0.06 cm}
2\pi\tau_2\kappa^{-1}((J^3)^2\hs{0.03 cm}+\hs{0.06 cm}(\bar J^3)^2)
\hs{0.04 cm}+\hs{0.04 cm}2\pi i\hs{0.02 cm}U_1(J^3-\bar J^3)}
\non\\
=\ C\hs{0.08 cm}q\bar q^{\hs{0.04 cm}-(c_{-\kappa}-1)/24}\hs{0.19 cm}
{\rm tr}|_{{\hs{-0.53 cm}{\rm level\hs{0.08 cm}}0}
\atop{m_l+m_r=0}}\hs{0.1 cm}q^{L_0^{\rm cs}}
\hs{0.09 cm}\bar q^{\bar L_0^{\rm cs}}\hs{0.08 cm}{\rm e}^{
\hs{0.03 cm}2\pi i\hs{0.03 cm}(UJ^3_0-\bar UJ^3_0)}\hs{0.8 cm}
\en
\no with the coset Virasoro generators
\be
L_0^{\rm cs}=L_0+\frac{_1}{^\kappa}\sum\limits_n:J^3_nJ^3_{-n}:\ ,\ \ \
\bar L_0^{\rm cs}=\bar L_0+\frac{_1}{^\kappa}\sum\limits_n:\bar J^3_n
\bar J^3_{-n}:\ .
\en
\no The contribution of the higher level
oscillatory modes is, however, less transparent than one
may naively think if we want to interpret
it in terms of gauge invariant states.
\vs 0.3 cm

\no\underline{4.4\ \ Asymmetric parafermions}
\vs 0.2 cm

Let us compare the situation to a somewhat similar
case of a variant of rational parafermionic theory which may be
described as the $SU(2)$ WZW model with the axial gauging of the $U(1)$
subgroup, i.e. with the diagonal $U(1)$ gauged asymmetrically.
The twisted toroidal partition function for such parafermions is
\cite{gepnqiu},\cite{coset}
\be
{\cal Z}^{\hs{0.02 cm}\rm pf}(U,\tau)\ =\ \int{\rm e}^{\hs{0.03 cm}
-kS(\gamma_U^{}g\hs{0.02 cm}\gamma_U^\D,\hs{0.03 cm}
A)}\hs{0.07 cm}Dg\hs{0.05 cm}DA\ .
\en
\no The integration is now over real $A$. Parametrizing
$A$ as before by the Hodge decomposition, one arrives at
the formula
\be
{\cal Z}^{\hs{0.02 cm}\rm pf}(U,\tau)\
=\ C\tau_2^{-1/2}\hs{-0.1 cm}\int\limits_{\bC/({\bf Z}+\tau
{\bf Z})}\hs{-0.1 cm}{\rm e}^{\hs{0.03 cm}\pi k
\tau_2^{-1}(U_1-iu_2)^2}\hs{0.05 cm}{\cal Z}^{SU(2)}
(\tau,u)\hs{0.12 cm}
|\eta(\tau)|^2\hs{0.09 cm}d^2u
\en
\no where ${\cal Z}^{SU(2)}(\tau,u)$ is the asymmetrically
twisted partition function
of the rational $SU(2)$ WZW model:
\be
{\cal Z}^{SU(2)}(\tau,u)
\ =\ q\bar q^{\hs{0.04 cm}
-c_{k}/24}\ {\rm tr}\hs{0.18 cm}
q^{L_0}\bar q^{\bar L_0}\hs{0.03 cm}{\rm e}^{\hs{0.03 cm}
2\hs{-0.03 cm}\pi i\hs{0.04 cm}(uJ^3_0+\bar uJ^3_0)}\ .
\en
\no The trace is taken over the space of states
\be
\hat{\cal H}^{SU(2)}\ =\ {_{_\bigoplus}\atop^{^{j\leq k/2}}}
\hat{\cal H}_j\otimes\overline{\hat{\cal H}}_j
\en
\no where $\hat{\cal H}_j$ carries the irreducible
spin $j$ level $k$ representation of the Kac-Moody
algebra $\hat{sl}(2,\bC)$. Notice the sign in
front of $\bar u\bar J^3_0$ in (25).
The integrand on the right
hand side of eq. (24) is a function on $\bC/({\bf Z}+\tau{\bf Z})$
only if $U_1\in k^{-1}{\bf Z}$ and only such twists should be
allowed. For other twists there is a global gauge anomaly:
the ungauged global $U(1)$ symmetry is broken in the parafermionic
theory to ${\bf Z}_k$. The spaces $\hat{\cal H}_j$ may be decomposed
into the weight spaces according to the integral or half-integral
eigenvalue $m$ of $J^3_0$ and at the same time with respect to the
level $k$ representations of the $\hat U(1)$ affine algebra
(similarly for the complex conjugates):
\be
\hat{\cal H}_j\cong{{_\bigoplus}\atop{^{^m}}}\hat{\cal H}_{j,m}^{\rm sing}
\otimes\hat{\cal H}'_m
\en
\no where $\hat{\cal H}^{\rm sing}_{j,m}$ is the subspace of $\hat{\cal H}_j$
where $J^3_0=m$ and $J^3_n=0$ for $n>0$. ${\cal H}'_m$ is the space
of the level $k$ $J^3_0=m$ irreducible representation
of the $\hat U(1)$ algebra.
The Sugawara Virasoro generator $L_0$ decomposes into the sum of
$L^{\rm cs}_0\equiv L_0-\frac{_1}{^k}\sum_n$:$J^3_nJ^3_{-n}$:
acting on spaces $\hat{\cal H}^{\rm sing}_{j,m}$ and $L'_0\equiv
\frac{_1}{^k}\sum_n$:$J^3_nJ^3_{-n}$: acting on $\hat{\cal H}'_m$
(in fact on $\hat{\cal H}^{\rm sing}_{j,m}$, $L^{\rm cs}_0=L_0-
\frac{_1}{^\kappa}m^2$).
Accordingly, we obtain for the partition function of the $SU(2)$
WZW theory:
\be
{\cal Z}^{SU(2)}(\tau,u)\ =\ (q\bar q)^{\hs{0.03 cm}
-(c_{k}-1)/24}\hs{0.08 cm}\sum\limits_{m_l,m_r}
{\cal Z}_{m_l,m_r}^{\rm sing}\hs{0.17 cm}q^{\hs{0.03 cm}
m_l^2/k}
\hs{0.04 cm}\bar q^{\hs{0.03 cm}m_r^2/k}\hs{0.1 cm}
|\eta(\tau)|^{-2}\hs{0.04 cm}
{\rm e}^{\hs{0.03 cm}2\pi i\hs{0.03 cm}(u\hs{0.03 cm}m_l+\bar u
\hs{0.03 cm}m_r)}
\en
\no where
\be
{\cal Z}_{m_l,m_r}^{\rm sing}\ =\
{\rm tr}|_{\hat{\cal H}^{\rm sing}_{m_l,m_r}}
\hs{0.17 cm}q^{L^{\rm cs}_0}\hs{0.04 cm}\bar q^{\bar L^{\rm cs}_0}
\en
\no with
\be
\hat{\cal H}^{\rm sing}_{m_l,m_r}\ \equiv\ {{_\bigoplus}\atop{^{^j}}}
\hat{\cal H}_{j,m_l}^{\rm sing}\otimes\overline{
\hat{\cal H}}_{j,m_r}^{\rm sing}\ .
\en
\no ${\cal Z}^{\rm sing}_{m_l,m_r}$ depends only on
$m_l$ and $m_r$ mod $k/2$ \cite{gepnqiu} (essentially due to the compact
nature of the gauged symmetry). More exactly,
\be
{\rm tr}|_{\hat{\cal H}^{\rm sing}_{j,m}}\hs{0.1 cm}q^{L_0^{\rm cs}}=
{\rm tr}|_{\hat{\cal H}^{\rm sing}_{j,m+k}}\hs{0.1 cm}q^{L_0^{\rm cs}}=
{\rm tr}|_{\hat{\cal H}^{\rm sing}_{k-j,-m}}\hs{0.1 cm}q^{L_0^{\rm cs}}\ ,\non
\en
\no see \cite{gepnqiu}. Upon the insertion of (28) into the
right hand side of (24), the $u_1$-integral will enforce equality
$m_l=-m_r\equiv m$. The sum over $m$ may be reduced mod $k/2$, with the
sum over the integral part of $2m/k$ used to extend the
integration over $u_2$ to a gaussian one over the entire real
line. Finally we get
\be
{\cal Z}^{\hs{0.03 cm}{\rm pf}}(\tau,U)\ =\
C\hs{0.05 cm}q\bar q^{\hs{0.05 cm}-(c_{k}-1)/24}\hs{0.08 cm}
\sum\limits_{m=0,\frac{_1}{^2},...,\frac{_k}{^2}}\hs{0.03 cm}
{\cal Z}^{\rm sing}_{m,-m}\hs{0.13 cm}{\rm e}^{\hs{0.03 cm}
-4\pi i\hs{0.03 cm}mU_1}\ .
\en
\vs 0.2 cm

As we see, the parafermionic partition function is
consistent (modulo multiplicity)
with the space of states of the coset theory
obtained by imposing the gauge conditions
\be
J^3_0+\bar J^3_0=0,\ \ J^3_n=\bar J^3_n=0{\rm\ \ for}\ \ n>0
\en
\no in the space of states of the ungauged WZW theory
with the Virasoro algebra given by the coset construction.
On the other hand, we could replace the first gauge condition
by $J^3_0+\bar J^3=kn$ for $n\in{\bf Z}$ or by $J^3=-\bar J^3$
and obtain equivalent theory. The latter means that the asymmetric
parafermions are indistinguishable from the symmetric ones.
\vs 0.3 cm

\no\underline{4.5\ \ Space of states}
\vs 0.2 cm

The level zero contribution (21) to the black hole
partition function is fully consistent with the gauge conditions
(32) imposed on states of the $H^+_3$ WZW theory
(for zero modes, only the first condition of
(32) restricts the states). The problem appears on the excited
levels of the space of states $\hat{\cal H}^{H^+_3}$ of the
ungauged theory. Let us consider, as an example, the first
excited level with states of the form
\be
\sum\limits_{a=\pm,3}(J^a_{-1}\psi_a+\bar J^a_{-1}\bar\psi_a)
\en
\no where $\psi_a,\hs{0.1 cm}\bar\psi_a$ are level zero states, i.e.
functions on $H^+_3$. The $J^3_0+\bar J^3_0=0$ condition
translates into
\be
(J^3_0+\bar J^3_0\pm 1)\psi_\pm=0\ ,\ \ \
(J^3_0+\bar J^3_0\pm 1)\bar\psi_\pm=0\ ,\ \ \
(J^3_0+\bar J^3_0)\psi_3=0\ .
\en
\no The other conditions of (32) give
\be
\psi_3=\frac{_2}{^\kappa}(J^+_0\psi_+-J^-_0\psi_-)\ ,\ \ \
\bar\psi_3=\frac{_2}{^\kappa}(\bar J^+_0\bar\psi_+-\bar J^-_0
\bar\psi_-)\ .
\en
\no Notice, however, that in $L^2(H^+_3)$, $J^3_0+\bar J^3_0$
is antihermitian so it has imaginary spectrum. Thus non-trivial
solutions of (34) and (35) are not only out of $L^2(H^+_3)$ but
do not belong to the generalized eigenspaces of $J^3_0,
\hs{0.1 cm}\bar J^3_0$ (they have ${\rm e}^{\hs{0.03 cm}\pm\phi}$
dependence on $\phi$). At best, we have to change the Hilbert space.
Notice how the situation here differs from the case of
parafermions where no such problems arise.
We may understand the above difficulty
also by looking at the level one contribution to
the partion function (10) which involves integrals
\be
\tau_2^{-1}\int{\rm e}^{\hs{0.05 cm}-\pi\kappa\tau_2^{-1}
(U_1-u_1)^2-\pi(\kappa-2)\tau_2^{-1}u_2^2}
\hs{0.1 cm}|\sin(\pi u)|^{-2}\hs{0.1 cm}{\rm e}^{\hs{0.03 cm}
\pm 2\pi iu}\hs{0.15 cm}d^2u\non\\
=\ C\tau_2^{-1/2}\int{\rm e}^{\hs{0.05 cm}4\pi\tau_2
(\kappa-2)^{-1}(\Delta+1/4)}\hs{0.04 cm}(2\pi u_2,
{\rm e}^{-2\pi iu_1}v\hs{0.05 cm};\hs{0.05 cm}0,v)\ \ \non\\
\cdot\hs{0.14 cm}{\rm e}^{\hs{0.03 cm}
\pm 2\pi(iu_1-u_2)}\hs{0.12 cm}{\rm e}^{\hs{0.04 cm}
-\pi\kappa\tau_2^{-1}
(U_1-u_1)^2}\hs{0.12 cm}d^2u\hs{0.07 cm}d^2v\ .\ \
\en
\no By spectral analysis, we may decompose operators
${\rm e}^{\hs{0.03 cm}t\Delta}$ into the heat kernels
acting in the
generalized eigenspaces of $J^3_0,\hs{0.08 cm}\bar J^3_0$:
\be
{\rm e}^{\hs{0.03 cm}t\hs{0.09 cm}
(\Delta+1/4)}\hs{0.04 cm}(2\pi u_2,
{\rm e}^{-2\pi iu_1}v'\hs{0.05 cm};\hs{0.05 cm}0,v)
\ =\ \sum\limits_n\int{\cal K}_{n,\omega}(t;|v'|,|v|)\hs{0.2 cm}
{\rm e}^{\hs{0.03 cm}2\pi inu_1
-2\pi i\omega u_2}\hs{0.1 cm}d\omega\ .
\en
\no This allows to rewrite integrals (36) as
\be
C\hs{0.1 cm}
\sum\limits_{n}\hs{0.1 cm}
{\cal K}_{n,\mp i}(4\pi\tau_2
\kappa^{-1};|v|,|v|)\hs{0.2 cm}
{\rm e}^{\hs{0.03 cm}-\pi\tau_2\kappa^{-1}\hs{-0.04 cm}
(n\pm 1)^2\hs{0.03 cm}+\hs{0.03 cm}2\pi i(n\pm 1)U_1}
\hs{0.18 cm}d|v|^2
\en
\no involving the analytic continuation of heat kernels
${\cal K}_{n,\omega}$ to imaginary values of $\omega$.
The question is whether such an analytic continuation
(which exists) corresponds to a heat kernel in a different
Hilbert space.
\vs 0.2 cm

\underline{Summarizing}. the gauge conditions (32) do not
determine
unambiguously the space of states.
We have to supplement them
with regularity conditions specifying domains of the operators
that they involve (the same applies to the BRST definition
of gauge invariant states). Ultimately, we should be able
to build a Hilbert space of states at each level and to compute
the contribution to the partition function as a
trace of a heat kernel in such a space. We shall discuss
a candidate solution of this problem in Sec. 5.
\vs 0.2 cm

On top of the above difficulties with the interpretation
of the partition function (but not unrelated to them)
comes the fact that, as it stands, the
integral on the right hand side of eq. (10) diverges.
The source of this divergence is, as in the mini-space
approximation, the infinite volume of the target space.
This may be regularized for example by defining
\be
\tilde{\cal Z}^{\rm bh}_{\rm reg}(\tau,U;R)\ =\ C\tau_2^{-1}
\int{\rm e}^{\hs{0.03 cm}-\pi\kappa\tau_2^{-1}|U-u|^2}
\hs{0.1 cm}{\cal S}(\tau,u)
\left(1-{\rm e}^{\hs{0.03 cm}R^2\hs{0.04 cm}{\cal S}(\tau,u)^{-1}}
\right)d^2u
\en
\no where
\be
{\cal S}(\tau,u)\ \equiv\ q\bar q^{\hs{0.05 cm}-1/12}\hs{0.16 cm}
{\rm e}^{\hs{0.03 cm}2\pi\tau_2^{-1}u_2^2}\hs{0.1 cm}|{\rm sin}
(\pi u)|^{-2}\bigg|\prod\limits_{n=1}^\infty(1-{\rm e}^{2\pi iu}
q^n)(1-{\rm e}^{\hs{0.03 cm}-2\pi iu}q^n)\bigg|^{-2}\ .
\en
\no The partition function $\tilde{\cal Z}^{\rm bh}_{\rm reg}
(\tau,U)$ is finite and when $R\rightarrow\infty$ and for $U_2=0$,
we recover the infinite integral (10) (we have put the twists
along both homology cycles in
$\tilde{\cal Z}^{\rm bh}_{\rm reg}(\tau,U)$ so that in the limit
$R\rightarrow\infty$ it corresponds to the black hole functional
integral with boundary conditions $v(z+2\pi)={\rm e}^{\hs{0.03 cm}
-2\pi i\Phi}v(z)$, $v(z+2\pi\tau)={\rm e}^{\hs{0.03 cm}
-2\pi i\Theta}v(z)$ where $U=\Theta-\tau\Phi$); for $\Phi=0$,
we recover $Z^{\rm bh}(\tau,U)$ with twist only along one cycle).
${\cal S}(\tau,U)$ is invariant under translations $U\longmapsto
U+n+\tau m$ for $n,m$ integers and is modular invariant.
As a result, under $SL(2,{\bf Z})$ transformations,
\be
\tilde{\cal Z}^{\rm bh}_{\rm reg}
(\frac{_{a\tau+b}}{^{c\tau+d}},\frac{_U}{^{c\tau+d}};R)
\ =\ \tilde{\cal Z}^{\rm bh}_{\rm reg}(\tau,U;R)\ ,
\en
\no i.e. the regularized partition function is modular covariant.
Again the divergence is logarithmic in $R$ and we could subtract
it to define the renormalized partition function measuring the
difference between the theories with the black hole and
(half-)cylinder targets.
\vs 0.3 cm

\no\underline{4.6\ \ Partition functions at higher genera}
\vs 0.2 cm

On a higher genus Riemann surface $\Sigma$ with the homology basis
$(a_\alpha,b_\beta)$, $\alpha,\beta=1,...,$genus, and with
the basic holomorphic
forms $\omega^\alpha$, $\int\limits_{a_\alpha}\omega^\beta=
\delta^{\alpha\beta}$,\ $\int\limits_{b_\alpha}
\omega^\beta=\tau^{\alpha\beta}\equiv\tau_1^{\alpha\beta}
+i\tau_2^{\alpha\beta}$, let us define the multivalued field
\be
\tilde\gamma_U(P)={\rm e}^{\hs{0.03 cm}\pi\sigma^3\hs{-0.1 cm}\int
\limits_{P_0}^{P}(U^{\rm t}\tau_2^{-1}\bar\omega-\bar U^{\rm t}
\tau_2^{-1}\omega)/2}
\en
\no with values in the Cartan subgroup of $SU(2)$.
Along the basic cycles
\be
\tilde\gamma_U(a_\alpha P)={\rm e}^{\hs{0.03 cm}-\pi
i\Phi_\alpha\sigma^3}.
\tilde\gamma_U(P)\ ,\non\\
\tilde\gamma_U(b_\alpha P)={\rm e}^{\hs{0.03 cm}-\pi
i\Theta_\alpha\sigma^3}
\tilde\gamma_U(P)\ \hs{0.17 cm}\non
\en
\no where $U=\Theta-\tau\Phi$.
The twisted partition function on $\Sigma$ is given by
\be
\tilde{\cal Z}^{bh}(\tau,U)\ =\ \int{\rm e}^{\hs{0.03 cm}\kappa
S(\tilde\gamma_U
\tilde\gamma_U^\D,\hs{0.03 cm}(2i)^{-1}\hs{-0.03 cm}A)}
\hs{0.14 cm}D(hh^\D)\hs{0.11 cm}DA
\en
\no with
\be
S(\tilde\gamma_Uhh^\D
\tilde\gamma_U^\D,\hs{0.03 cm}(2i)^{-1}\hs{-0.03 cm}A)\ =\
S(hh^\D,\hs{0.04 cm}\frac{_{1}}{^{2i}}(A+\pi\bar U^{\rm t}
\tau_2^{-1}\omega+\pi U^{\rm t}\tau_2^{-1}\bar\omega))\non\\
+\hs{0.1 cm}\frac{_1}{^{2i}}\int A\wedge(\bar U^{\rm t}\tau_2^{-1}
\omega-U^{\rm t}\tau_2^{-1}\bar\omega)\hs{0.1 cm}+\hs{0.1 cm}
\pi U^{\rm t}\tau_2^{-1}U\ .
\en
\no It defines the higher genus partition function for the black hole
with twists of the $v$-field by ${\rm e}^{\hs{0.03 cm}-2\pi i\Phi_\alpha}$
facto
the $a_\alpha$ cycles and by ${\rm e}^{\hs{0.03 cm}-2\pi i\Theta_\beta}$
along the $b_\beta$ ones. We decompose again the gauge field according to
Hodge:
\be
A=d\mu+*d\nu+\pi\bar u^{\rm t}\tau_2^{-1}\omega+\pi u^{\rm t}
\tau_2^{-1}\bar\omega
\en
\no and integrate over the $v$-field (of $hh^\D$), $\nu$ and $\mu$
(the latter integral gives the volume of the gauge group).
What is left is the $\phi$ functional integral and the integral
over twists $u$:
\be
\tilde{\cal Z}^{\rm bh}(\tau,U)\ =\ C\left({{_{{\rm det}'
(-\bar\p^{^*}\hs{-0.03 cm}
\bar\p)}}\over{^{\rm area}}}\right)^{1/2}\int{\rm e}^{\hs{0.03 cm}
-\pi\kappa(\bar U-\bar u)^{\rm t}\tau_2^{-1}(U-u)\hs{0.05 cm}+\hs{0.05 cm}
(2\pi i)^{-1}\hs{-0.06 cm}\kappa\hs{-0.04 cm}\int(\p\phi)(\bar\p\phi)}\non\\
\cdot\hs{0.14 cm}{\rm det}\left(\bar\p+\bar\p\phi+\pi u^{\rm t}\tau_2^{-1}
\bar\omega)^{^*}(\bar\p+\bar\p\phi+\pi u^{\rm t}\tau_2^{-1}
\bar\omega)\right)^{-1}\hs{0.01 cm}\delta(\phi(\xi_0))\hs{0.13 cm}
D\phi\hs{0.13 cm}d^{2{\hs{0.04 cm}\rm genus}}u\ .
\en
\no By the chiral anomaly (compare the genus one formula (3.8)),
\be
{\rm det}\hs{-0.04 cm}\left(\bar\p+\bar\p\phi+\pi u^{\rm t}\tau_2^{-1}
\bar\omega)^{^*}(\bar\p+\bar\p\phi+\pi u^{\rm t}\tau_2^{-1}
\bar\omega)\right)^{-1}\
=\ {\rm e}^{\hs{0.03 cm}i\pi^{-1}\hs{-0.06 cm}\int(\p{\phi})
(\bar{\p}\phi)
\hs{0.08 cm}
+\hs{0.05 cm}(2\pi i)^{-1}\hs{-0.08 cm}\int{\phi}{\cal R}}\non\\
\cdot\hs{0.2 cm}\left({\rm det}_{\alpha,\beta}
(\smallint{\rm e}^{\hs{0.03 cm}2\phi}\overline{\eta_{u\alpha}}{\eta_{u\beta}})
\hs{0.1 cm}/\hs{0.1 cm}{\rm det}_{\alpha,\beta}
(\smallint\overline{\eta_{u\alpha}}{\eta_{u\beta}})
\right)^{-1}\hs{0.06 cm}{\rm det}\hs{-0.04 cm}
\left(\bar\p_{u}^{^*}\hs{0.01 cm}\bar\p_{u}\right)^{-1}
\en
\no where $\bar\p_{u}\equiv\bar\p+\pi u^{\rm t}\tau_2^{-1}
\bar\omega$ and $\eta_{u\alpha},\ \alpha=1,...,{\rm genus}-1,$
form a basis of the 01-forms in the kernel of $\bar\p_{u}^{^*}$. Using
eq. (47), we may rewrite the partition function as
\be
\tilde{\cal Z}^{\rm bh}(\tau,U)\ =\ C\left({{_{{\rm det}'
(-\bar\p^{^*}\hs{-0.03 cm}\bar\p)}}\over{^{\rm area}}}\right)^{1/2}\int{\rm
e}^{
-\pi\kappa(\bar U-\bar u)^{\rm t}\tau_2^{-1}(U-u)\hs{0.05 cm}+\hs{0.05 cm}
(2\pi i)^{-1}\hs{-0.06 cm}(\kappa-2)\hs{-0.04 cm}
\int(\p\phi)(\bar\p\phi)\hs{0.07 cm}+\hs{0.04 cm}(2\pi i)^{-1}
\hs{-0.1 cm}\int\hs{-0.05 cm}{\phi}{\cal R}}\non\\
\cdot\hs{0.14 cm}{\rm e}^{\hs{0.03 cm}-\hs{-0.06 cm}\int\hs{-0.06 cm}\bar
\eta_{u}\exp(2\phi)\hs{0.05 cm}\eta_{u}}\hs{0.1 cm}{\rm det}\hs{-0.04 cm}
\left(\bar\p_{u}^{^*}\hs{0.01 cm}\bar\p_{u}\right)^{\hs{-0.04 cm}-1}
\hs{-0.04 cm}
\delta(\phi(\xi_0))\hs{0.1 cm}D\phi\hs{0.1 cm}
d\eta_{u}\hs{0.1 cm}du\ \ \
\en
\no where the gaussian integral over $\eta_{u}\in{\rm ker}\hs{0.07 cm}
\p_{u}^{^*}$ was used to express the $\eta_{u\alpha}$ determinants.
The expression is obviously similar to the Liouville partition function
although the real relation between two theories lies probably
deeper. In any way, we expect the $\phi$ and $\eta$ integrals
to be finite and to lead to an expression regular in $u$ except
for the contribution of $\hs{0.07 cm}
{\rm det}\left(\bar\p_{u}^{^*}\hs{0.01 cm}\bar
\p_{u}\right)^{\hs{-0.04 cm}-1}$ which around $u=0$
behaves as $|u|^{-2}$ which is integrable for genus$\ >1$
and diverges logarithmically for genus 1 ($\eta_{u\alpha}$ may be
chosen regular in $u$ around $u=0$). This singularity
is repeated around other points of ${\bf Z}+\tau{\bf Z}$.
Thus, similarly as for the Liouville theory coupled to free
bosonic field, see \cite{david},\cite{distkaw},
we expect
the partition functions at higher genera to be convergent
reflecting the finite dimension of the region in
the target space relevant for the stringy interaction.
\vs 0.3 cm

\no\underline{4.7\ \ Green functions}
\vs 0.2 cm

Since the coset theory is an instance of a gauge theory,
its Green functions should be given by functional integral with
gauge invariant insertions.
Examples of gauge invariant fields are $f_{\rho,m_l,m_r}(v(\xi))$
of eq. (3.33) with $m_l=-m_r\equiv m$ whose conformal weights are
\be
\Delta_{\rho,m}=\bar\Delta_{\rho,m}=\frac{_{1+\rho^2}}
{^{4(\kappa-2)}}+\frac{_{m^2}}{^{\kappa}}\ .
\en
\no If we instead used $f_{\rho,m_l,m_r}(\phi(\xi),v(\xi))$
with $m_l+m_r\not=0$ as local
fields, we could still maintain local gauge invariance by adding
compensating currents, i.e. by considering insertions
\be
I(hh^\D,\frac{_1}{^{2i}}A)\
=\ \prod\limits_\alpha f_{\rho_\alpha,m_{l\alpha},
m_{r\alpha}}(\phi(\xi_\alpha),v(\xi_\alpha))
\hs{0.18 cm}{\rm e}^{\hs{0.03 cm}
-\int\limits_{^c}A}
\en
\no where $c$ is a chain such that
$\delta c=\sum\limits_\alpha (m_{l\alpha}+m_{r\alpha})\hs{0.03 cm}
\xi_\alpha$.
In the planar case, the functional integral over the gauge
field may be easily done upon parametrization $A=d\mu+*d\nu$.
The integral over $\mu$ drops out because of gauge invariance
and the integral over $\nu$ gives expectation value of chiral
vertex operators
\be
\int{\rm e}^{\hs{0.1 cm}i\hs{-0.1 cm}\int\limits_{^{^{c+c'}}}
\hs{-0.1 cm}\p\nu\hs{0.05 cm}
-\hs{0.05 cm}i\hs{-0.1 cm}\int\limits_{^{^{c-c'}}}\hs{-0.1 cm}
\bar\p\nu\hs{0.05 cm}-\hs{0.05 cm}\pi^{-1}\hs{-0.03 cm}\kappa\hs{-0.05 cm}
\int(\p_z\nu)\hs{0.03 cm}
(\p_{\bar z}\nu)\hs{0.04 cm}d^2z}\hs{0.1 cm}D\nu
\en
\no where $\delta c'=\sum\limits_\alpha (m_{l\alpha}-m_{r\alpha})
\xi_\alpha$ (compare \cite{coset} where similar calculation was done for
the parafermions). Altogether, we obtain
\be
\int I(hh^\D,\frac{_1}{^{2i}}A)\ {\rm e}^{\hs{0.03 cm}\kappa
\hs{0.03 cm}S(hh^\D,\hs{0.08cm}(2i)^{-1}\hs{-0.07 cm}A)}
\hs{0.08 cm}D(hh^\D)\hs{0.14 cm}DA\hs{4 cm}\non\\
=\ {\rm const}.\ \prod\limits_{\alpha\not=\alpha'}(\xi_\alpha-
\xi_{\alpha'})^{m_{l\alpha} m_{l{\alpha'}}
/\kappa}\hs{0.12 cm}(\bar\xi_\alpha-
\bar\xi_{\alpha'})^{m_{r\alpha}m_{r{\alpha'}}/\kappa}
\hs{0.24 cm}\int I(hh^\D,0)\ {\rm e}^{\hs{0.03 cm}\kappa
\hs{0.01 cm}S(hh^\D)}
\hs{0.07 cm}D(hh^\D)
\en
\no where
the $\prod\limits_{\alpha\not=\alpha'}$ factors
come from the (properly renormalized) free field integral (51).
They modify the conformal dimensions of fields
$f_{\rho,m_l,m_r}$ of the $H^+_3$ WZW theory to
\be
\Delta_{\rho,m_l}=\frac{_{1+\rho^2}}{^{4(\kappa-2)}}+\frac{_{m_l^2}}
{^{\kappa}}\ ,\ \ \ \bar\Delta_{\rho,m_r}=\frac{_{1+\rho^2}}
{^{4(\kappa-2)}}+\frac{_{m_r^2}}{^{\kappa}}\ .
\en
\no producing operators with imaginary spin and hence never
local. It is possible, however, that correlations of fields
coming from common eigenfunctions on $H^+_3$ of $\Delta,J^3,\bar J^3$
which do not correspond to the spectrum, for example for $\omega$
imaginary, may be given sense. If in the left hand side of (52)
we integrated out
the $A$-field, we would obtain the black hole functional integral
with insertions which for large values of
$|v(\xi_\alpha)|$ take form
\be
\prod\limits_\alpha\left(\hs{0.07 cm}|v(\xi_\alpha)|^{m_{l\alpha}+m_{r_\alpha}}
\hs{0.06 cm}f_{\rho_\alpha,m_{l\alpha},m_{r\alpha}}
(0,|v(\xi_\alpha|)\hs{0.07 cm}\right)
\hs{0.08 cm}{\rm e}^{\hs{0.03 cm}-i\hs{-0.08 cm}\int\limits_{^{^{c+c'}}}
\hs{-0.09 cm}\p\hs{0.03 cm}
{\rm arg}\hs{0.03 cm}(v)\hs{0.05 cm}+\hs{0.05 cm}i\hs{-0.08 cm}
\int\limits_{^{^{c-c'}}}
\hs{-0.09 cm}\bar\p\hs{0.03 cm}{\rm arg}\hs{0.03 cm}(v)}\ .
\en
\no We recover then the chiral vertex operators of field
${\rm arg}\hs{0.03 cm}(v)(\xi)$ which, for large $|v|$, becomes
a compactified free field. If fields with real $m_l+m_r$ existed, they
would be mutually local for $m_l+m_r\in\kappa{\bf Z}$, as are their
asymptotic versions. We shall return to the discussion of this possibility
in the next section.
\vs 0.5 cm

\no 5.\ \ $SU(1,1)$ mod $U(1)$ COSET THEORY
\vs 0.3 cm
\no\underline{5.1\ \ Functional integral formulation}
\vs 0.2 cm
\addtocounter{equation}{-54}

The original proposal \cite{w3} for the conformal sigma model with
2D black hole target was based on a coset construction
starting with $SU(1,1)\cong SL(2,{\bf R})$ WZW model.
The parametrization
\be
g\ =\ \left(\matrix{{\rm e}^{\hs{0.03 cm}i\psi}(1+|v|^2)^{1/2}&v\cr
\bar v&{\rm e}^{-i\psi}(1+|v|^2)^{1/2}\cr}\right)
\en
\no where $\psi$ is in ${\bf R}/(2\pi{\bf Z})$ and $v$ is complex
gives  global coordinates on $SU(1,1)$. Comparing
to parametrization (3.1) of positive elements in $SL(2,\bC)$, we see
that it passes to the present one by simple substitution
$\phi\mapsto i\psi$.
Consequently, for the WZW action with the $U(1)\subset SU(1,1)$
gauged asymmetrically (i.e. with the axial $U(1)$ gauge), we obtain
from eq. (3.4)
\be
S(g,\frac{_1}{^{2i}}A)\ =\ -\frac{_1}{^{\pi}}
\int[\hs{0.05 cm}
(i\p_z\tilde\psi+A_z)(i\p_{\bar z}\tilde\psi+A_{\bar z})
\non\\
+\hs{0.05 cm}(\p_z+i\p_z\tilde\psi+A_z)\bar v\hs{0.08 cm}
(\p_{\bar z}+i\p_{\bar z}\tilde\psi+A_{\bar z})v\hs{0.05 cm}]
\hs{0.08 cm}d^2z\ .
\en
\no where $\tilde\psi\equiv\psi+\frac{_1}{^2}i\hs{0.04 cm}
{\rm log}(1+|v|^2)$. The axial gauge invariance is
\be
S({\rm e}^{\hs{0.03 cm}i\lambda\sigma^3}g\hs{0.05 cm}
{\rm e}^{\hs{0.03 cm}i\lambda\sigma^3},\hs{0.03 cm}\frac{_1}{^{2i}}
(A-2id\lambda)\hs{0.04 cm})\ .
\en
\no The euclidean action
$\pm\kappa\hs{0.03 cm}S(g)$ for the $SU(1,1)$ WZW theory is
not bounded below. For the minus sign (and $\kappa$ positive)
this is due to the $\tilde\psi$-field contribution. As a result,
the stable euclidean picture is missing for this theory.
In the coset functional integral
\be
\int\ -\ {\rm e}^{\hs{0.03 cm}\kappa\hs{0.01 cm}S(g,
\hs{0.03 cm}(2i)^{-1}\hs{-0.07 cm}A)}\hs{0.13 cm}Dg\hs{0.1 cm}DA\ ,\non
\en
\no however, the $\tilde\psi$-field may
be gauged out and absorbed by a translation of $A$. If
$A$ is taken real then the $A$ integral is stable
and the translation of $A$ is complex (the axial gauge
invariance requires imaginary $A$). In this case, moreover,
after the translation, we recover the same integral as before
for the $SU(2,\bC)/SU(2)$ mod ${\bf R}$ coset theory.
It seems that the two coset theories coincide\footnote{this is the
point on which the present author's opinion has wavered most
and might continue to do so with the progress in the understanding
of both theories}. On the quantum-mechanical level, the equivalence
of both approaches may be seen clearly.
\vs 0.3 cm

\no\underline{5.2\ \ Particle on SU(1,1)}
\vs 0.2 cm

The classical
mini-space system which corresponds to the 2D WZW theory
with target $SU(1,1)$ is the geodesic motion in the invariant
metric on $SU(1,1)$ of signature, say, $(-,+,+)$. We
may quantize it taking $L^2(SU(1,1))$ with the Haar measure
(equal $d\psi\hs{0.04 cm}d^2v$ in parametrization (1))
as the space of states in which
$SU(1,1)_{\rm left}\times SU(1,1)_{\rm right}$ acts unitarily.
Infinitesimally, we get the action of $sl(2,\bC)\oplus sl(2,\bC)$
generated by $J^a$'s and $\bar J^a$'s given by the same
formulae as in the case of $L^2(H^+_3)$ except for the substitution
$\phi\mapsto i\psi$. The hermiticity relations change, however,
and we obtain
\be
{J^a}^*=-J^{a}\ ,\ \ \ {\bar J}^{a^*}=-\bar J^a\ \ \ {\rm for}\ a=1,2\ ,\non\\
{J^3}^*=J^3\ ,\ \ \ \ {\bar J}^{3^*}=\bar J^3\ \hs{3.04 cm}.
\en
\no Also $-\Delta\equiv-{\vec J}^{2\atop{}}=-{\vec{\bar J}}^{_2}$ is no more
bounded below. It is again given explicitly by eq. (3.12) with
$\p_\phi^2$ replaced by $-\p_\psi^2$. In fact
\be
L^2(SU(1,1))\
\cong\ \int\limits_{{\rho>0}\atop{\epsilon=0,1/2}}
\hs{-0.41 cm}^{^{^{^\bigoplus}}}
\hs{0.09 cm}{\cal D}_{\rho,\hs{0.03 cm}\epsilon}\otimes
\bar{\cal D}_{\rho,\hs{0.04 cm}\epsilon}\hs{0.16 cm}
d\nu(\epsilon,\sigma)\hs{0.3 cm}\bigoplus
\bigoplus\limits_{{j=-1,-3/2,...}\atop{\pm}}
{\cal D}^\pm_{j}\otimes\bar{\cal D}^\pm_{j}\ .
\en
\no ${\cal D}_{\sigma,\epsilon}$ carry
unitary irreducible representations of $SU(1,1)$ of
the principal continuous series which may be realized
in the space of sections of a spin bundle on the circle
($SU(1,1)$ acts naturally on $S^1$, $\epsilon$ corresponds
to two choices of the spin structure). The eigenvalue
of $\vec{J}^{2\atop{}}$ on ${\cal D}_{\rho,\epsilon}$ is
$-\frac{_1}{^4}(1+\rho^2)$.
Spaces ${\cal D}^\pm_{-j}$ carry the lowest- (highest)-weight
representations of $sl(2,\bC)$ of spin $j$. They give the discrete
series of unitary, irreducible representations of $SU(1,1)$
with eigenvalue of $\vec{J}^{2\atop{}}$ equal to $j(j+1)$ which is $\geq 0$.
If, instead of $SU(1,1)$, we considered its simply-connected
covering $\widetilde{SU(1,1)}$ (where $\psi$ takes values in the
non-compactified real line), the direct sums in decomposition (5) over
$\epsilon$ and $j$ would be replaced by direct integrals
over $0\leq\epsilon<1$ and $j<-1/2$.
Since $-\Delta$ plays the role of Hamiltonian, the
energy is not bounded below (nor above). This problem with
stability renders the above quantization physically not very
satisfactory. Indeed, the way we proceeded here is not the one
used for example to quantize a particle in Minkowski space
where one recovers satisfactory solution of the stability
problem passing to the second-quantized level. Finding
a stable quantization of the particle on $SU(1,1)$ or,
more importantly, of the $SU(1,1)$ WZW field theory remains
an open and seemingly very interesting problem\footnote{we
thank G. Gibbons for attracting our attention to it}. Here,
however, we shall be interested only in the coset
$SU(1,1)$ mod $U(1)$ theory where coupling to the gauge
field removes the unstable $\tilde\psi$ field. On the
quantum-mechanical level, the gauge condition $J^3+\bar J^3
=i\p_\psi=0$, cuts out from $L^2(SU(1,1))$ the contribution
of the discrete series (and more) making $-\Delta$ positive.
Besides,
\be
L^2(SU(1,1))|_{J^3+\bar J^3=0}\cong L^2(d^2v)
\cong L^2(H^+_3)|_{J^3+\bar J^3=0}\non
\en
\vs 0.1 cm
\no in a natural way and this isomorphism preserves
(restrictions of) $\Delta,\ J^3\ $and $\bar J^3$.
This proves on the mini-space level the identity of the coset
theories $SU(1,1)$ mod $U(1)$ and $SL(2,\bC)/SU(2)$
mod ${\bf R}$. The generalized eigenfunctions $f_{\rho,m_l,m_r}$
of $\Delta,\hs{0.1 cm}J^3,\hs{0.1 cm}\bar J^3$ on $SU(1,1)$,
corresponding to eigenvalues $-\frac{_1}{^4}(1+\rho^2),\hs{0.05 cm}m_l,
\hs{0.05 cm}m_r$ with $m_l\pm m_r\in{\bf Z}$, are
given by Jacobi functions \cite{vilen}. For $m_l+m_r=0$ they are independent
of $\psi$ and, although given by different expressions,
coincide with similar eigenfunctions on $H^+_3$.
For example, from the harmonic analysis on $SU(1,1)$, we obtain
\be
f_{0,0,0}(v)\ =\ \pi\int\limits_0^{2\pi}(1+2|v|^2+2
v(1+|v|^2)^{1/2}\hs{0.03 cm}{\rm cos}
\hs{0.04 cm}\theta\hs{0.03 cm})^{-1/2}\hs{0.05 cm}d\theta
\en
\no which should be compared with eq. (3.34).
For both $m_l+m_r$ equal and different from zero, eigenfunctions
$f_{\rho,m_l,m_r}$ seem to generate primary fields of dimensions
given by eq. (4.53) (for $m_l+m_r\not=0$, they should be dressed
with line integrals of the gauge field, like in (4.50)). If
$m_l+m_r\in\kappa{\bf Z}$, the corresponding fields are
mutually local.
\vs 0.38 cm

\no\underline{5.3\ \ Space of states, unitarity, duality, problems}
\vs 0.28 cm

On the level of 2D field theories,
neither $SL(2,\bC)/SU(2)$ mod $\bf R$ nor $SU(1,1)$ mod $U(1)$
theory has been shown to exist, least solved completely, so comparison
is more difficult. The computation of the partition function
in the first case did not require complex rotations or shifts
of the fields so it seems more trustable. Nevertheless,
we have seen that the interpretation of the excited
contributions to it
required analytic continuation of the heat kernels
on the eigensubspaces of $J^3,\hs{0.1 cm}\bar J^3$
in $L^2(H^+_3)$ to imaginary eigenvalues $\omega$
of $\frac{_1}{^i}(J^3+\bar J^3)=i\p_\phi$. But this
should be given by the heat kernels in the eigenspaces
of $J^3,\hs{0.1 cm}\bar J^3$ in $L^2({SU(1,1)})$, or more generally
in $L^2(\widetilde{SU(1,1)})$\hs{0.09 cm}, obtained
by the substitution $\phi\mapsto i\psi$.
It is then possible that the partition function becomes a trace
over the gauge-invariant states of the $SU(1,1)$ WZW theory.
Superficially, the $U(1)$ coset of the latter has the same
problem as the parafermionic model discussed in Sec. 4.4: the
ungauged (vector) $U(1)$
symmetry has global anomaly which seems to break $U(1)$ to
${\bf Z}_k$. Here, this is a spurious problem, however: if we
start from the $\widetilde{SU(1,1)}$ WZW theory rather than from
the $SU(1,1)$ one, the coset theory is the same but the
complete $U(1)$ symmetry is present.
The space of states of the $\widetilde{SU(1,1)}$ WZW theory
should be a subspace of
\be
\int\limits_{{\rho>0}\atop{0\leq\epsilon<1}}
\hs{-0.41 cm}^{^{^{^\bigoplus}}}
\hs{0.09 cm}\hat{{\cal D}}_{\rho,\hs{0.03 cm}\epsilon}\otimes
\hat{\bar{\cal D}}_{\rho,\hs{0.04 cm}\epsilon}\hs{0.16 cm}
d\nu(\epsilon,\sigma)\hs{0.3 cm}\bigoplus
\int\limits_{{j<-1/2}\atop{\pm}}
\hs{-0.41 cm}^{^{^{^\bigoplus}}}
\hs{0.09 cm}\hat{{\cal D}}^\pm_{j}
\otimes\hat{\bar{\cal D}}^\pm_{j}
\en
\no where ``$\hs{0.07 cm}\hat{\ }\hs{0.07 cm}$'' denotes
the representation space of
the Kac-Moody algebra $\hat{sl}(2,\bC)$ induced (in the sense of
Sec. 3.3) from the representations of $\widetilde{SU(1,1)}$.
What exactly should be the subspace taken does not seem
to be clear yet. A possibility is the appearance of the fusion
rule $-\frac{_1}{^2}(\kappa-1)<j$ in the
discrete series, analogous to the
rule $j\leq k/2$ of the $SU(2)$ WZW theory.
Spaces $\hat{\cal D}$ may be provided with the hermitian form
for which ${J^a_n}^*=-J^a_{-n}$ for $a=1,2$ and ${J^3_n}^*=J^3_{-n}$
(this agrees at level zero with the scalar product induced from
$L^2(SU(1,1))$\hs{0.1 cm}).
The encouraging sign is the important result of Dixon-Lykken-Peskin
\cite{dlp} (see also \cite{lzuck}) who proved that the gauge conditions
$J^3_n=0,\ n>0,$ cut out, under the
restriction $-\frac{_1}{^2}\kappa\leq j$ on the discrete series,
the negative norm states from the induced representations.
Notice, that the latter condition disposes of the representations
with negative eigenvalues of $L^{\rm cs}_0$. Their absence
should then be assured by stability if the coincidence with
the explicitly stable $H^+_3$ mod $\bf R$ model really takes
place. In that case, the $SU(1,1)$ mod $U(1)$ approach should
allow to show the unitarity of the euclidean black hole CFT.
Moreover, we should be able to assemble the calculated partition
functions from the characters of the induced representations
$\hat{\cal D}$. This is not simple even on the quantum
mechanical level where we know that it works.
\vs 0.25 cm

The gauge condition $J^3_0+\bar J^3_0=0$ leaves us with spin-less,
$U(1)$-charge zero sector of the theory. The functional
integral for the partition function, as in any gauge theory,
should be given by the trace over this subspace of states,
as is also clearly indicated by the $U$ dependence of the result
(4.10). The primary fields $f_{\rho,m_l,m_r}$ with $m_l+m_r=0$ correspond
to vectors in this sector. On the other hand, gauge conditions
$J^3_0+\bar J^3_0=l\kappa$ should give for $0\not=l\in{\bf Z}$ sectors
with the spin and the $U(1)$ charge different from zero.
Fields $f_{\rho,m_l,m_r}$
with $m_l+m_r=l\kappa$ should correspond to states in these
sectors. From the point of view of
the asymptotic free field
with the cylindrical part of the cigar as the target, these
are the winding sectors, see formula (4.54)\footnote{one should
not confuse these sectors with
the infinite volume superselection sectors obtained by sending
some of the charges ``behind the moon''; they are rather descendents
of the charge sectors corresponding to infinitely heavy external
charges - we thank E. Seiler for correcting some of the author's
original misconceptions about this point}.  The partition functions
corresponding to the winding sectors can be also computed, essentially
by inserting a Polyakov line with charge $l\kappa$
into the functional integral. We plan to return to these issues
elsewhere.
\vs 0.2 cm

Another open problem in the black hole CFT
is a relation between the coset models
$SU(1,1)$ mod $U(1)$ obtained by gauging the axial and the vector $U(1)$
subgroup. The vector theory has a more serious stability problem than
the  axial one since the vector gauging does not seem to remove
completely the unbounded below modes. On a rather formal level
one can argue that both theories have the same spectrum of mutually
local operators \cite{dvv}-\cite{kirit},\cite{rocver}.
It was expected that they give the same CFT.
The vector coset results in a sigma model with singular metric on
the target. In the asymptotic
region, the target also looks like a half-cylinder and the identity
of the models would become there that of free fields compactified
on dual radia \cite{kadanoff}. We have not been able, however, to stabilize
the functional integral for the vector theory in a sensible way
to show that it has the same partition function
as the axial coset. The situation should be contrasted with the case
of parafermions. There, as we have seen in Sec. 4.4, both gaugings
give the same theory, in fact already on the mini-space level.
In particular, both partition functions coincide.
The duality between the two $U(1)$ cosets of the $SU(1,1)$ WZW
theory requires, in our opinion, further study. It may be that
the vector description may be maintained only in the asymptotically
flat region. The issue
is important for understanding whether the coupling to dynamical gravity
washes out the singularity at $v^+v^-=1$ of the classical
Minkowskian metric (4.5) interchanged by the duality with
the non-singular horizon $v^+v^-=0$, see \cite{dvv}.
Even less clear is what sense we can make of the sigma model
which Minkowskian 2D black hole as the target which formally
comes from gauging non-compact subgroup in $SU(1,1)$ theory
\cite{w3} and how all these theories fit together. We clearly
touch here on the relation between the stability
and unitarity of the CFT's and the signature of the effective
targets. If a progress can be made in understanding such issues
fundamental for quantum gravity,
the effort invested in studying relatively simple
non-rational theories may pay back.
\vs 0.6 cm

\end{document}